\documentclass[twocolumn, aps, pra, amsmath, amssymb, nofootinbib, superscriptaddress, longbibliography, floatfix, table-of-contents, eqsecnum]{revtex4-2}

\usepackage{nicematrix}
\usepackage{tikz}

\usepackage[]{graphicx}
\usepackage{mathrsfs}
\usepackage{times}

\usepackage[colorlinks = true, linkcolor=teal, citecolor=teal, urlcolor  = teal]{hyperref}

\usepackage{array}
\usepackage{comment}
\usepackage{mathtools}

\usepackage{type1cm}
\usepackage{lettrine}
\usepackage[english]{babel}
\usepackage{microtype}
\usepackage{booktabs}
\usepackage[T1]{fontenc}
\usepackage[boxed, vlined]{algorithm2e}
\usepackage[margin=0pt, font=small, labelfont=bf, labelsep=endash, 
justification=centerlast, 
labelsep=colon]{caption}

\usepackage{braket}
\usepackage[toc,page]{appendix}

\usepackage{amssymb}
\usepackage{mathrsfs}

\usepackage{multirow}
\usepackage{etoolbox}
\usepackage{breqn}
\usepackage{float}

\usepackage{amsthm}
\newtheorem{theorem}{Theorem}

\definecolor{darkgreen}{rgb}{0.0, 0.42, 0.24}

\newcommand\numberthis{\addtocounter{equation}{1}\tag{\theequation}}

\usepackage{mathtools}
\def\multiset#1#2{\ensuremath{\left(\kern-.3em\left(\genfrac{}{}{0pt}{}{#1}{#2}\right)\kern-.3em\right)}}

\makeatletter
\preto\maketitle{%
  \begingroup\lccode`~=`,
  \lowercase{\endgroup
  \let\saved@breqn@active@comma~
  \let~}\active@comma 
}
\appto\maketitle{%
  \begingroup\lccode`~=`,
  \lowercase{\endgroup
  \let~}\saved@breqn@active@comma 
}
\makeatother

\begin{document}

\title{Coherently mitigating boson samplers with stochastic errors}

\author{Deepesh Singh}
\email[]{deepesh.sang@gmail.com}
\affiliation{Centre for Quantum Computation and Communications Technology, School of Mathematics and Physics, The University of Queensland, Brisbane, Queensland 4072, Australia}

\author{Ryan J.\ Marshman}
\affiliation{Centre for Quantum Computation and Communications Technology, School of Mathematics and Physics, The University of Queensland, Brisbane, Queensland 4072, Australia}

\author{Nathan Walk}
\affiliation{Dahlem Center for Complex Quantum Systems, Freie Universität Berlin, 14195 Berlin, Germany}

\author{Jens Eisert}
\affiliation{Dahlem Center for Complex Quantum Systems, Freie Universität Berlin, 14195 Berlin, Germany}
\address{Helmholtz-Zentrum Berlin f{\"u}r Materialien und Energie, 14109 Berlin, Germany}

\author{Timothy C.\ Ralph}
\affiliation{Centre for Quantum Computation and Communications Technology, School of Mathematics and Physics, The University of Queensland, Brisbane, Queensland 4072, Australia}

\author{Austin P.\ Lund}
\affiliation{Dahlem Center for Complex Quantum Systems, Freie Universität Berlin, 14195 Berlin, Germany}
\affiliation{Centre for Quantum Computation and Communications Technology, School of Mathematics and Physics, The University of Queensland, Brisbane, Queensland 4072, Australia}

\begin{abstract}
Sampling experiments provide a viable route to show quantum advantages of quantum devices over classical computers in well-defined computational tasks. However, quantum devices such as boson samplers are susceptible to various errors, including stochastic errors due to fabrication imperfections. These cause the implemented unitary operations to deviate randomly from their intended targets, following distributions with finite variance. Whilst full-scale quantum error correction remains challenging in the near term, quantum error mitigation schemes have been devised to estimate expectation values, but it is unclear how these schemes would work for sampling experiments. In this work, we demonstrate that, given access to multiple stochastic unitaries, it is possible to mitigate the effect of these errors in sampling experiments.
We adopt the unitary averaging protocol 
which employs multiple stochastic boson samplers to generate a distribution that 
approximates the ideal boson sampler distribution as the number of samplers increases. 
We derive a
rigorous upper bound on the trace distance between the output probability distributions induced by invertible vacuum-heralded 
networks based on
the Schur-Weyl duality. 
This result can be seen concretely as an error mitigation scheme in sampling experiments against stochastic errors. On a broader level, it suggests a path towards understanding error mitigation for sampling experiments and developing analysis tools for photonic circuits incorporating measurements and feed-forward. 
We further provide other applications of unitary averaging, including its use 
in implementing the linear combination of unitaries and benchmarking fabrication repeatability in linear optics.
\end{abstract}

\maketitle


\section{Introduction} \label{intro}

Quantum information processing has emerged as a frontier in modern physics, offering the potential for groundbreaking advancements in computation, communication, and sensing. At the core of these quantum protocols is the manipulation of quantum states -- a process that depends critically on our ability to perform precise transformations on these states. Photonic systems, with their relative ease of implementation and well-understood properties, have long been considered a promising candidate for realising these quantum protocols \cite{Photonic,RevModPhys.79.135,Photonic2,Fusion,Xanadu,QuantumPhotoThermodynamics,TeleportationReview}.
%
However, while featuring substantial advantages in some respects, 
these systems face significant challenges in achieving precise control due to various sources of errors.

While loss and photon distinguishability are well-known and significant sources of error in photonic quantum computing, the precise control required to implement target unitary transformations presents an additional, often underappreciated challenge. The inherent limitations in accurately implementing state transformations in photonic experiments stem from fabrication imperfections, also described in Refs.\ \cite{Burgwal:17, Russell_2017}. As a result, any attempted unitary operation in a photonic system does not result in the precise implementation of the target unitary but rather a different quantum channel. Oftentimes -- in particular when fabrication errors come into play -- this 
quantum channel can be well approximated by a randomly selected unitary from a distribution centred around the target 
with some finite variance. We expect the specific noise distribution to depend on the experimental setup or fabrication process.

These stochastic errors, combined with the challenges of loss and photon distinguishability, pose significant obstacles to realising the full potential of quantum information processing using photonic platforms. Generating improved quantum states typically requires full error correction, which, although efficient in principle, is extremely challenging to implement in practice due to its high resource costs. Although various error mitigation techniques have been proposed to overcome these limitations \cite{RevModPhys.95.045005}, 
they are generally limited to the construction of good approximations of the expectation values of observables \cite{Li:2017,Temme:2017,Endo:2018,Cai:2023,RevModPhys.95.045005}. 

Our study builds and further develops an alternate protocol called unitary averaging \cite{PhysRevA.110.012457, ryan2018, ryan2024, nibedita2024} that is here seen to occupy a middle ground between error mitigation and error correction,
and that is generalised to cover sampling schemes as well. It involves coherent encoding and redundancy similar to quantum error correction or detection while requiring less overhead, reminiscent of quantum error mitigation. It shares similarities with coherent error 
mitigation techniques such as virtual distillation \cite{PhysRevX.11.041036} and exponential error suppression \cite{PhysRevX.11.031057}, 
quantum error mitigation with classical shadows
\cite{PRXQuantum.4.010303},
and to some extent, with mitigated 
readout schemes \cite{EmilioPaper}, as they all involve the coherent manipulation of quantum states comprising multiple copies of a quantum circuit.
This approach leverages multiple stochastic unitaries to probabilistically generate improved quantum states without correcting for all error syndromes. We demonstrate that it is possible to produce a state that converges to the ideal state as the number of redundant unitaries increases.

While, in principle, one may be able to correct for all error syndromes, we avoid this to maintain experimental feasibility. Our approach requires only passive linear optical elements, preventing the need for active optical elements or feed-forward mechanisms. This makes our protocol much easier to implement, offering a promising and practical solution to the challenges posed by stochastic errors on different photonic platforms, including those arising from imperfect unitary implementations in integrated photonics.

In this work, we provide an exact realisation of the unitary averaging 
framework in the context of boson samplers \cite{AaransonArkhipov},
instances of optical quantum random sampling schemes aimed at optically
achieving a quantum advantage over classical computers by making use of
sampling experiments \cite{SupremacyReview}. 
Given access to multiple stochastic boson samplers randomly picked from a distribution centred around the ideal boson sampler, we show that the unitary averaging framework can generate distributions that converge to the ideal boson sampling distribution as the number of redundant boson samplers increases. The unitary averaging protocol is inherently probabilistic, with a success probability that has a lower bound scaling exponentially with the depth of the boson sampler and the number of photons. This scaling can be compared with the similar scaling of large classes of 
mostly classical error mitigation schemes, for which it is known that the number of samples for mitigating possibly non-unital
local noise scales exponentially in the product of the number of qubits and depth
\cite{ErrorMitigationObstructionsOld,Quek:2024}
in the worst case. Despite these daunting worst-case bounds, it is still plausible that these intermediate schemes can have a practical benefit. In our case of $n$ photons propagating through a depth $d$ interferometer with a stochastic noise per beamsplitter $\nu$, even for a success probability that scales as $\exp(-\nu n d)$, in practice, $\nu$ can be very small such that for certain finite-size regimes of order $\sim 1/\nu$, a benefit from unitary averaging can be feasible (for similar arguments regarding sample complexity bounds on error mitigation see, e.g., Ref.\  \cite{Zimboras:2025}). Furthermore, the unitary averaging scheme outputs a bona fide quantum state, complete with any coherence and entanglement that may be desired both internally and with an external state. As such, it can be utilised within larger, more complex quantum computation schemes or as an improved input to a full-scale quantum error correction in a manner that no classical or incoherent/semi-classical scheme can.



To support our results, we derive an upper bound on the total variation distance between the output probability distributions induced by linear invertible transformations from vacuum-heralded linear optical networks using results from representation theory. This bound validates our protocol and has broader applications in quantum optics. The \emph{Knill-Laflamme-Milburn} protocol \cite{knill2001klm} has provided a paradigmatic framework for implementing non-linearities using photon heralding, 
at least in principle, 
giving rise to a substantial body of 
more resource-efficient methods for linear-optical quantum computing \cite{RevModPhys.79.135,PhysRevLett.95.010501,PhysRevLett.96.020501,Fusion, PhysRevA.74.042343}. However, a quantitative understanding of how much photon heralding is needed to simulate a given amount of non-linearity has still been lacking. Our methods shed light on the power of vacuum heralding and provide insights into the degree of non-linearity they can simulate. 

Our work also contributes to understanding error detection and correction in photonic quantum systems and highlights the potential of probabilistic approaches in potentially overcoming hardware limitations. 
Furthermore, the unitary averaging framework has direct connections 
to the topics of \emph{linear combination of unitaries} (LCU) \cite{10.5555/2481569.2481570} and stochastic error cancellation \cite{cai_et_al:LIPIcs.TQC.2024.2}, which are critical tools 
in quantum simulation 
and 
error mitigation. 

In the following sections, we first review the impact of imperfect unitary implementations on boson sampling distributions in Sec.\ \ref{background}. Next, we define our problem setup in Sec.\ \ref{problemSetup} and compare the unitary averaging protocol with the naive distribution averaging protocol in Sec.\ \ref{results}. In Sec.\ \ref{proof}, we derive an upper bound on the distance between output probability distributions generated by two vacuum-heralded linear interferometers. We then discuss our results in Sec.\ \ref{discussion} and the broader implications of our findings in Sec.\ \ref{furtherConsiderations}. Finally, we conclude our work in Sec.\ \ref{conclusion}.

\section{Background}
\label{background}
In the boson sampling experiment, unitary noise arises from imperfections in the optical components such as beam splitters, phase shifters, and mirrors. These imperfections lead to deviations from the ideal unitary transformation that defines the boson sampling problem. Instead of the intended unitary matrix \( U \), the actual transformation implemented can be represented as another unitary matrix \( V \), which closely approximates \( U \). We provide an explicit example of this error model in Sec.\ \ref{problemSetup}.

The target distribution \( \mathcal{D}_U \) in boson sampling is the probability distribution of detecting different photon configurations at the output ports, given an initial photon input state and an ideal unitary transformation \( U \). The presence of unitary noise alters the output distribution, denoted as \( \mathcal{D}_V \), which can differ significantly from the ideal distribution \( \mathcal{D}_U \).

For a boson sampling experiment with an $m$ mode unitary $U$ and $n$ single photons input, the probability of detecting the photons in the output modes in a particular output configuration \( \mathbf{x} = (x_1, x_2, \ldots, x_m) \) such that $\sum_{i=1}^{m} x_{i} = n$ is given by
\begin{equation} 
 P_U(\mathbf{x}) = \frac{ \left| \mathrm{Perm}(U_{\mathbf{x}}) \right|^2}{x_1 ! x_2 ! \ldots x_m!},
 \end{equation}
where \( \mathrm{Perm}(U_{\mathbf{x}}) \) denotes the permanent of a constructed matrix \( U_{\mathbf{x}} \), consisting of entries from \( U \) as described in Ref.~\cite{AaransonArkhipov}. In the presence of unitary noise, however, this probability transforms to
\begin{equation}
P_V(\mathbf{x}) = \frac{\left| \mathrm{Perm}(V_{\mathbf{x}}) \right|^2}{x_1 ! x_2 ! \ldots x_m!} 
\end{equation}
where \( V_{\mathbf{x}} \) is constructed from the noisy unitary matrix \( V \) in a similar manner to \(U_x\).

To quantify the impact of unitary noise on the complete boson sampling distribution and not just the probabilities of individual output photon configurations, we can compute the \emph{total variation distance} (TVD), also known as the classical trace distance, $L_1$ distance or Kolmogorov distance \cite{NielsenMichaelA}, between the ideal and noisy boson sampling distributions. The TVD between the boson sampling distributions $\mathcal{D}_U$ and $\mathcal{D}_{V}$ generated by unitaries $U$ and $V$, respectively, is defined as
\begin{equation} \Vert \mathcal{D}_U - \mathcal{D}_{V} \Vert = \frac{1}{2}  \sum_{\mathbf{x}} \left\vert \mathcal{D}_U(\mathbf{x}) - \mathcal{D}_{V}(\mathbf{x}) \right\vert, \end{equation}
where the sum is over all possible photon configurations $\mathbf{x}$. Clearly, a small TVD indicates that the noisy distribution closely approximates the ideal one.

Interestingly, the relationship between the TVD of the boson sampling distributions and the operator norm distance of the corresponding unitaries has been proven in Ref.\ \cite{Arkhipov_2015} to be
\begin{align}
    \Vert \mathcal{D}_{U} - \mathcal{D}_{V} \Vert \leq n \Vert U - V \Vert_{\mathrm{op}}.
    \label{eq:distance_bound}
\end{align}
This bound shows that the TVD between the ideal and noisy distributions is upper bounded by the operator norm distance between the ideal and noisy unitaries, scaled by the total number of photons $n$. Thus, even slight deviations in the unitary matrix can significantly change the output distribution, especially as the number of photons increases.

In summary, unitary noise significantly challenges the fidelity of the target distribution in boson sampling. Understanding and mitigating this noise is crucial for realising boson sampling's full potential as a tool for demonstrating quantum computational advantage.

\section{Problem Setup}
\label{problemSetup}
Although the results presented in this work can be used to analyse a broad class of vacuum-heralded protocols in linear optics, in this section, we explore the experimentally relevant problem involving stochastic unitary errors in boson samplers \cite{Burgwal:17, Russell_2017}. We assume that the stochastic unitaries are fixed over the sampling time, and hence, any fluctuations are negligible while generating the distribution. The essence of our problem lies in formulating a protocol to construct an output distribution using multiple copies of these noisy interferometers to minimise its distance from the one produced by the target unitary operation. 

We consider \(N\) linear interferometers implementing these boson samplers \cite{Reck, Clements, Russell_2017, Burgwal:17}, described by unitary matrices \(U_{1}, U_{2}, \dots, U_{N}\), where each \(U_i\) is an \emph{independent and identically distributed} 
(i.i.d.) stochastic realization of the target unitary \(U\).
For the sake of clarity and concreteness, we focus exclusively on classical stochastic errors. That said, as has been said elsewhere \cite{ClaraPaper}, suitably classical errors can closely resemble quantum noise captured by dynamical semi-groups generated by Lindbladians, so that one should expect
that the scheme presented here is also robust under
errors that are not genuinely classical in nature.

Each \(U_i\) is constructed using beam splitters \(B_{i,j}\), whose reflectivities and phases approximate the parameters of \(U\).
Using the notation used in Ref.\ \cite{Clements}, the beam splitter \(B_{i,j}\) for \(U_i\) can be expressed as
\begin{equation}
    B_{i,j} = \begin{bmatrix}
        e^{i \phi^{'}} \cos \theta^{'}_{i,j} & -\sin \theta^{'}_{i,j} \\
        e^{i \phi^{'}} \sin \theta^{'}_{i,j} & \cos \theta^{'}_{i,j}
    \end{bmatrix}, \label{eq:U_definition}
\end{equation}
where the noisy parameters are given by
\begin{equation}
\theta^{'}_{i,j} = \theta_{i,j} + \delta \theta_{i,j}, \quad \text{and} \quad
\phi^{'}_{i,j} = \phi_{i,j} + \delta \phi_{i,j}.
\end{equation}
Here, \(\theta_{i,j}\) represents the target beam splitter reflectivity, and \(\phi_{i,j}\) represents the target phase-shift. The noise terms \(\delta \theta_{i,j}, \text{ and } \delta \phi_{i,j}\) are small stochastic perturbations such that 
\begin{equation}
\delta \theta_{i,j}, \delta \phi_{i,j} \ll 1.
\end{equation}
Without loss of generality, assuming the noise parameters are i.i.d.\  with a Gaussian noise profile, each noise term \(\delta x_{i,j} \in \{\delta \theta_{i,j}, \delta \phi_{i,j}\}\) satisfies the statistical properties 
\begin{align}
\mathbb{E}( \delta x_{i,j}) &= 0, \nonumber \\
\mathbb{E} (\delta x_{i,j} \delta x_{p,q} ) &= \nu \delta_{i,p} \delta_{j,q},  \nonumber \\
\mathbb{E} (\delta x_{i,j}^3 ) &= 0, \nonumber \quad  \\
\mathbb{E} (\delta x_{i,j}^4 ) &= 3\nu^2,
\label{eq:noise_stats}
\end{align}
with parameter variance $\nu$.
Using decomposition methods such as those by Reck \cite{Reck} and Clements \cite{Clements}, any element of the unitary matrix \(U_i\) can be written as a product of the entries from its constituent beam splitters \(B_{i,j}\). Thus, \(U_i\) is a stochastic realization of an approximation of the target unitary \(U\), with 
matrix elements expressed as
\begin{equation}
    \sum \left(\prod\limits_{\mathbf{r}} \cos \theta^{'}_{\mathbf{r}} \prod\limits_{\mathbf{s}} \sin \theta^{'}_{\mathbf{s}} \prod\limits_{\mathbf{t}} e^{i \phi^{'}_{\mathbf{t}}} \right),
\end{equation}
where $\mathbf{r}, \mathbf{s}, \text{ and } \mathbf{t}$, together determine the different paths that a photon follows in the interferometer. 

Since each beam splitter's parameters are independent, the mean value of each matrix element of \(U_i\) is given by
\begin{equation}
    \sum \left(\prod\limits_{\mathbf{r}} \mathbb{E}(\cos  \theta^{'}_{\mathbf{r}}) \prod\limits_{\mathbf{s}} \mathbb{E}( \sin \theta^{'}_{\mathbf{s}}) \prod\limits_{\mathbf{t}} \mathbb{E} (e^{i \phi^{'}_{\mathbf{t}}}) \right).
\end{equation}
Using the Taylor expansion to second order and the statistical properties of the error terms from Eq.\ \eqref{eq:noise_stats}, we find that for \(y \in \{\cos \theta^{'}_{\mathbf{r}}, \sin \theta^{'}_{\mathbf{s}}, e^{i \phi^{'}_{\mathbf{t}}}\}\), the approximation
\begin{align}
     \mathbb{E}(y)  \approx \Big(1 - \frac{\nu}{2}\Big) y_0
\end{align}
holds,
where \(y_0 \in \{\cos \theta_{\mathbf{r}}, \sin \theta_{\mathbf{s}}, e^{i \phi_{\mathbf{t}}}\}\) are the target values, respectively.

In general, an \(m\)-mode optical interferometer can be implemented with a depth \(d = m\) using \(m(m-1)/2\) variable beam splitters, as described in Clements' decomposition \cite{Clements}. To ensure a uniform depth, additional phase shifters can be introduced in the interferometer, such that all optical paths traverse exactly \(d\) beamsplitter and phase-shifter elements. While this uniform-depth configuration simplifies analytical calculations, it comes at the cost of introducing additional noise into the system.

We adopt this uniform-depth model to facilitate the derivation of analytical expressions for the success probability of the unitary averaging protocol discussed in Section~\ref{UAProtocol}. However, the success probability for the uniform-depth model, denoted as \(p_{\mathrm{uni}}\), is a conservative estimate and provides a lower bound on the actual success probability \(p_{\mathrm{post}}\), achievable with the original Clements' decomposition.

Under the uniform-depth assumption, where \(|\mathbf{r}| + |\mathbf{s}| + |\mathbf{t}| = d\) for all paths, the mean value of each element of the average unitary matrix is given by
\begin{equation}
     \Big(1 - \frac{\nu}{2}\Big)^d \sum \left(\prod\limits_{\mathbf{r}} \cos \theta_{\mathbf{r}} \prod\limits_{\mathbf{s}} \sin \theta_{\mathbf{s}} \prod\limits_{\mathbf{t}} e^{i \phi_{\mathbf{t}}} \right).
\end{equation}
Consequently, for all noisy unitaries \(U_i\), the relationship between the mean value of \(U_i\) and the target unitary \(U\) can be expressed as
\begin{equation}
     \mathbb{E}(U_i) = \Big(1 - \frac{\nu}{2}\Big)^d U. \label{eq:3.9}
\end{equation}

\section{Main results}
\label{results}

This section compares two protocols for generating distributions that approximate the target boson sampling distribution. The first, known as the \textit{distribution averaging protocol}, involves generating \(N\) individual boson sampling distributions from \(N\) stochastic unitaries and averaging these distributions in post-processing to estimate the target boson sampling distribution. The second, the \textit{unitary averaging protocol}, first averages the \(N\) stochastic unitaries coherently and probabilistically to construct an effective averaged unitary, generating a normalised distribution as the estimator of the target boson sampling distribution.

We analyse the performance of both approaches by examining the distance between the target distribution and the corresponding estimator distribution, and we numerically demonstrate that the unitary averaging protocol offers an improvement over the distribution averaging framework at the cost of a smaller success probability. 

\begin{figure}[htbp]
    \centering
 \includegraphics[width=0.42\textwidth]{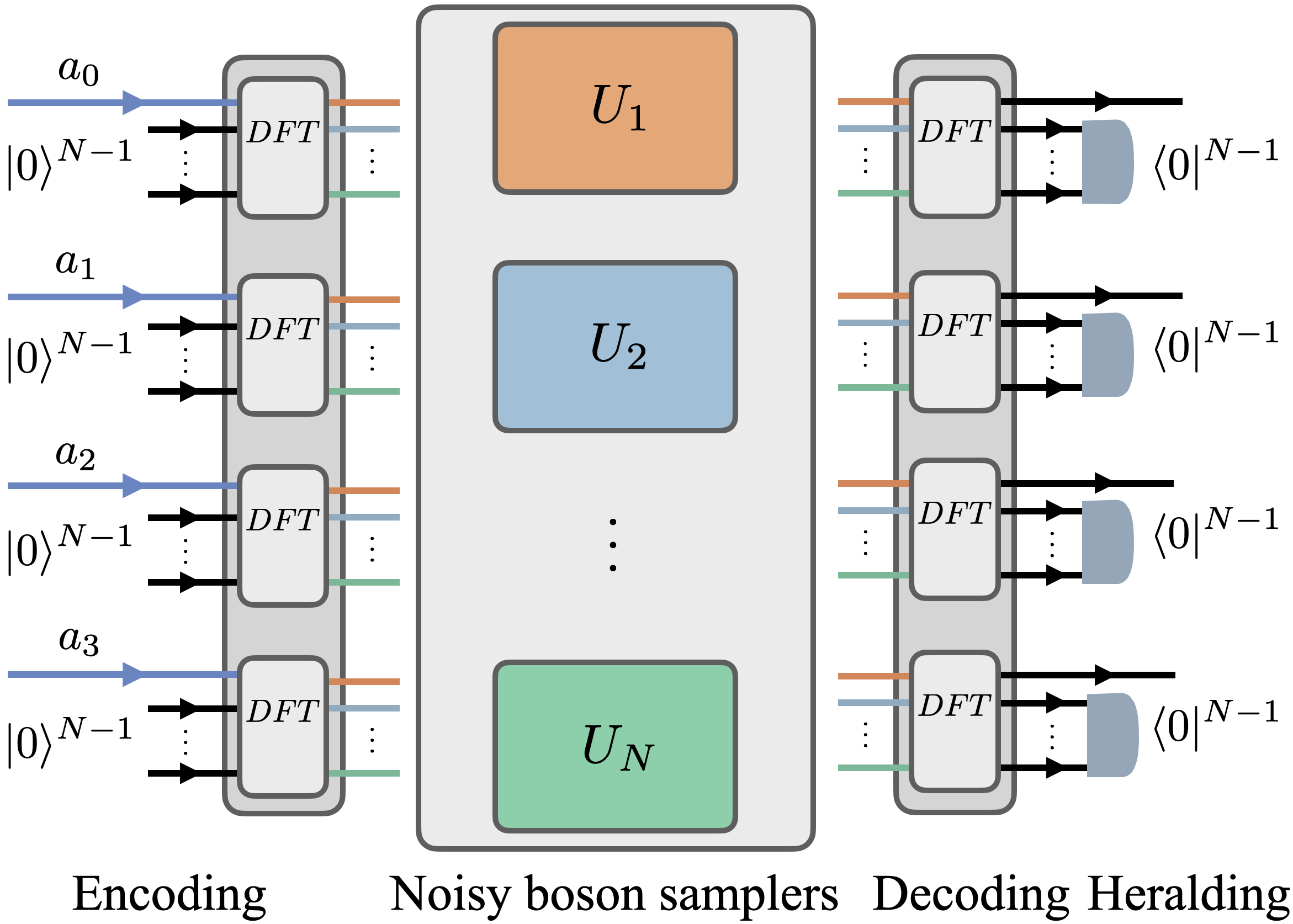}
    \caption{Implementation of an averaged unitary action on $4$ input modes. Each original input mode is encoded with $N - 1$ vacuum modes using $N$-dimensional DFT operators. The setup uses $N$ redundant copies of the $m$-mode unitary \( U_i \), where \( 1 \leq i \leq N \), followed by decoding with DFTs and vacuum heralding on the auxiliary modes. The modes are colour-coded to indicate which unitary they belong to, as described in Ref.\ \cite{PhysRevA.110.012457}.
    }
    \label{fig:averaged_unitary}
\end{figure}

\subsection{Distribution averaging protocol}

The distribution averaging protocol serves as a naive error mitigation strategy by averaging the \(N\) boson sampling distributions generated from \(N\) stochastic boson samplers. We analyze the distance \(\Vert \mathcal{D}_{U} - \mathbb{E} (\mathcal{D}_{U_i}) \Vert\) between the target boson sampling distribution \(\mathcal{D}_{U}\) and the averaged distribution \( \mathbb{E} (\mathcal{D}_{U_i} )\), where \(\mathcal{D}_{U_i}\) represents the distributions produced by the \(N\) stochastic unitaries \(U_{1}, U_{2}, \dots, U_{N}\), and \(\mathbb{E} (\mathcal{D}_{U_i} )\) is the average of these distributions. Using the triangle inequality, we get 
\begin{align}
    \Vert \mathcal{D}_{U} - \mathbb{E} (\mathcal{D}_{U_i}) \Vert \leq \mathbb{E} \left( \Vert \mathcal{D}_{U} - \mathcal{D}_{U_i} \Vert \right),
\end{align}
which can be simplified further using Eq.~\eqref{eq:distance_bound} as
\begin{align}
    \mathbb{E} \left( \Vert \mathcal{D}_{U} - \mathcal{D}_{U_i} \Vert \right) \leq n \mathbb{E} \left( \Vert U - U_i \Vert_{\mathrm{op}} \right).
\end{align}

The error model in Eq.~\eqref{eq:U_definition} allows us to express each noisy unitary as \(U_i = U \delta U_i\), where \(U\) is the target unitary and \(\delta U_i\) represents the error. Due to the invariance of the operator norm under unitary multiplication, 
\begin{align}
    \mathbb{E} \left( \Vert \mathcal{D}_{U} - \mathcal{D}_{U_i} \Vert \right) & \leq n \mathbb{E} \left( \Vert U (I - \delta U_i) \Vert_{\mathrm{op}} \right) \nonumber \\
    & \leq n \mathbb{E} \left( \Vert I - \delta U_i \Vert_{\mathrm{op}} \right).
\end{align}
Since \(I - \delta U_i \geq 0\) in general, we have
\begin{equation}
    0 \leq \mathbb{E} \left( \Vert I - \delta U_i \Vert_{\mathrm{op}} \right) \leq c, 
\end{equation}
for some constant $c \geq 0$. Therefore,
\begin{align}
    \mathbb{E} \left( \Vert \mathcal{D}_{U} - \mathcal{D}_{U_i} \Vert \right) \leq cn,
\end{align}
indicating that, in general, the distribution averaging protocol does not guarantee convergence to the target distribution.

\subsection{Unitary averaging protocol}
\label{UAProtocol}

The \emph{unitary averaging}  (UA) framework is a probabilistic protocol that works by converting stochastic errors into heralded loss. UA allows one to get higher quality states by utilising the redundancy of the noisy transformations and vacuum heralding. The Hadamard encoding, vacuum ancilla, auxiliary modes, and the on/off detectors employed in this passive protocol are readily available in linear optics, making it the most suitable architecture for realising this framework. An example implementation of the unitary averaging protocol is shown in Fig.~\ref{fig:averaged_unitary}.

Given access to multiple noisy unitaries, the UA framework allows one to apply a coherent average of these unitaries on the intended modes, with the success probability of this transformation depending on the exact value of the unitaries and the state on which the transformation is applied. It should be noted that the UA protocol has an additional, qualitative advantage over error mitigation, in the sense that, given access to multiple noisy unitaries, 
it provides an error-suppressed state instead of just the expectation values of observables, which can be utilised in further computation. Furthermore, it is a less resource-intensive protocol than error correction since no active elements and corrections for arbitrary syndrome measurements are involved. Therefore, UA defines a new class of error-suppression protocols that sit somewhere between error-mitigation and error-correction protocols, probabilistically converting stochastic errors in networks to heralded loss. 

Specifically, UA acts by applying an averaged unitary evolution given by
\begin{equation}
	U_{\mathrm{avg}}
    \coloneqq\frac{1}{N}\sum_{j=1}^{N}U_j
\end{equation}
on a target set of $M$ modes using $N$ noisy copies of a target unitary $U$, each labelled $U_{j}$. An accompanying $(N-1)M$ set of error modes must be heralded in the vacuum state. If each applied unitary $U_j$ is approximately implementing a target unitary $U$ with unbiased, i.i.d.\  noise, then UA will apply an averaged unitary $U_{\mathrm{avg}}$. This will be a stochastic operator approximating the target unitary $U$, with variance reduced by a factor of $N$ when compared to each of the original transformations $U_j$.

The central limit theorem applies to the individual matrix elements of the averaged transformation, decreasing variances as \( 1/N \), where \( N \) is the number of redundant encodings. The exact form of the average matrix and the distribution of transformations over a finite number of averages depend on the specifics of the noise affecting the network encoding. 

\begin{figure}[htbp]
    \centering
    \includegraphics[width=0.48\textwidth]{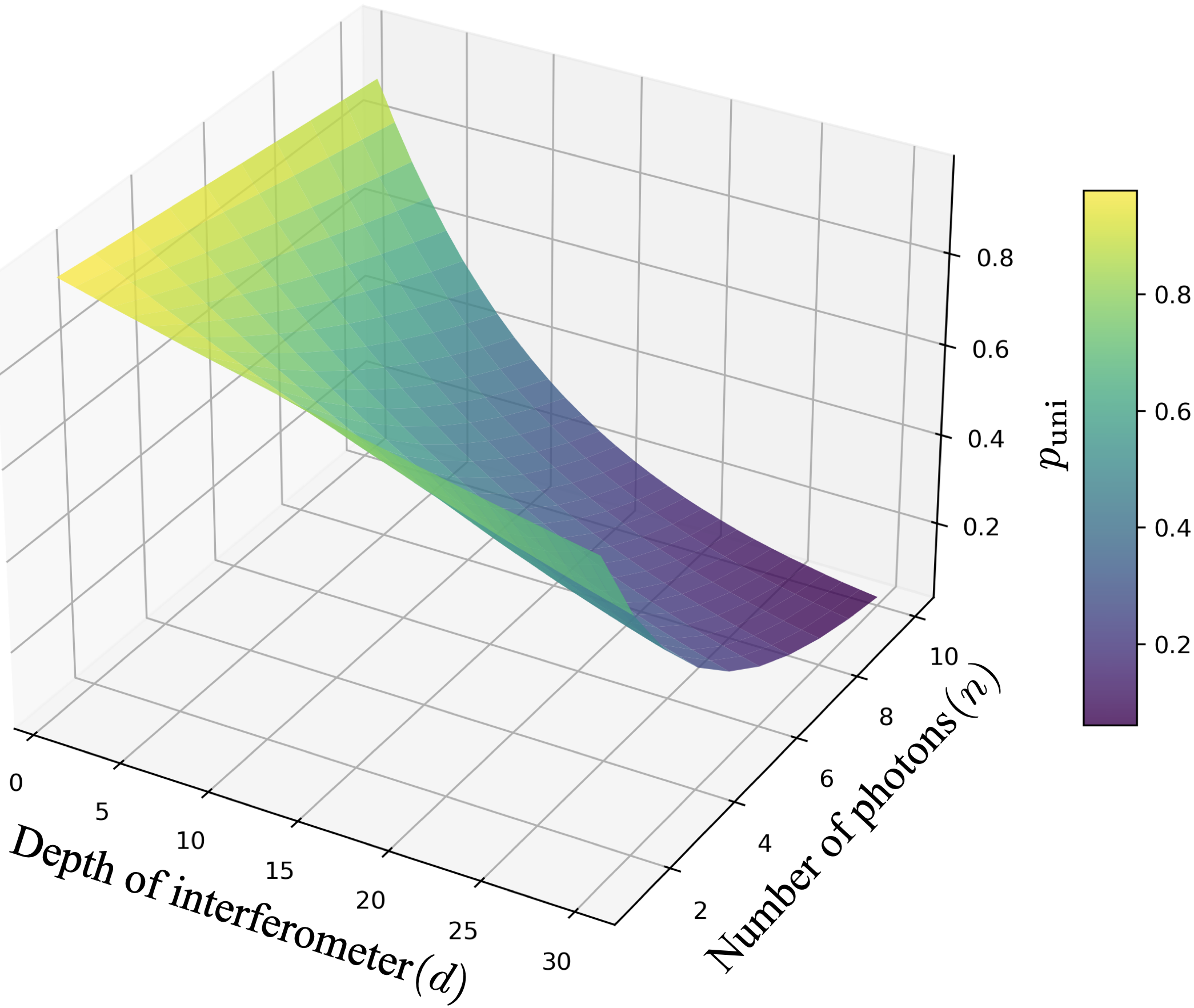}
    \caption{The success probability of the unitary averaging protocol in the limit $N \rightarrow \infty$ under the uniform depth assumption ($p_{\mathrm{uni}}$) is shown as a function of the depth ($d$) and the number of photons ($n$) in a boson sampler. The variance of each tunable element of the interferometer implementing the boson sampler is fixed at $\nu = 0.01$. Notably, $p_{\mathrm{uni}}$ serves as a lower bound for the true success probability, $p_{\mathrm{post}}$, of the unitary averaging protocol in the limit $N \rightarrow \infty$ when using the actual Clements decomposition with non-uniform depth.
}
    \label{fig:p_uni_plot}
\end{figure}

However, analysing the UA protocol becomes easier in the infinite averaging ($N \rightarrow \infty$) case. Since in this limit, the transformation $U_{\mathrm{avg}}$ can be approximated by the mean value of $U_i$, i.e., 
$\mathbb{E}(U_{i})$, we can write the output state vector $\vert \psi_{\mathrm{out}} \rangle$ of the UA framework as
\begin{equation}
    \vert \psi_{\mathrm{out}} \rangle = \frac{\phi \left( \mathbb{E} (U_{i}) \right)}{\sqrt{p_{\mathrm{post}}}} \vert \psi_{\mathrm{in}} \rangle, \label{eq:outstate}
\end{equation}
where $\vert \psi_{\mathrm{in}} \rangle$ is the input state vector to UA setup, $\phi$ is the homomorphism from a unitary acting on one boson to that acting on $n$ identical bosons \cite{AaransonArkhipov, Arkhipov_2015, Lund2023}, and $p_{\mathrm{post}}$ is the success probability of the unitary averaging protocol defined in 
Sec.\ \ref{problemSetup}. Using the uniform depth implementation defined in Sec. \ref{problemSetup}, we have a depth $d$ and success probability $p_{\mathrm{uni}}$. We can then use Eq. \eqref{eq:3.9} to rewrite the output state vector as
\begin{align}
     \vert \psi_{\mathrm{out}} \rangle = \Big(1 - \frac{\nu}{2} \Big)^{dn} \frac{\phi( U )}{\sqrt{p_{\mathrm{uni}}}} \vert \psi_{\mathrm{in}} \rangle. \label{eq:finalstate}
\end{align}
From Sec.\ \ref{problemSetup}, we also know that $p_{\mathrm{uni}} \leq p_{\mathrm{post}}$ and using the normalization condition, we get
\begin{equation}
    p_{\mathrm{uni}} = \Big(1 - \frac{\nu}{2} \Big)^{2dn}.
\end{equation}
Therefore, the lower bound for the success probability of the unitary averaging protocol in the infinite redundant encoding limit ($N \rightarrow \infty $) decreases exponentially in the depth of the boson sampling interferometer $d$ and the number of photons $n$, as shown in Fig.\  \ref{fig:p_uni_plot}. 

The average of the unitaries corresponding to noisy boson samplers, \( U_{\text{avg}} \), is generally non-unitary and represents an \( m \times m \) complex submatrix of the larger unitary interferometer defined by the encoding, redundant noisy boson samplers, and decoding unitaries. This 
implies \( U_{\text{avg}} \in \mathbb{C}^{m \times m} \) and satisfies \( \| U_{\text{avg}} \| \leq 1 \). Deriving an upper bound for the distance between distributions induced by two such averaged boson sampler setups requires new tools, as the distance bounds established in Ref.~\cite{Arkhipov_2015} are valid only for comparing boson sampling distributions produced by two unitary interferometers and do not extend to vacuum-heralded unitary interferometers, of which unitary averaging is a specific instance.

\begin{figure}[!htb]
    \centering{ 
 {\includegraphics[width=0.45\textwidth]{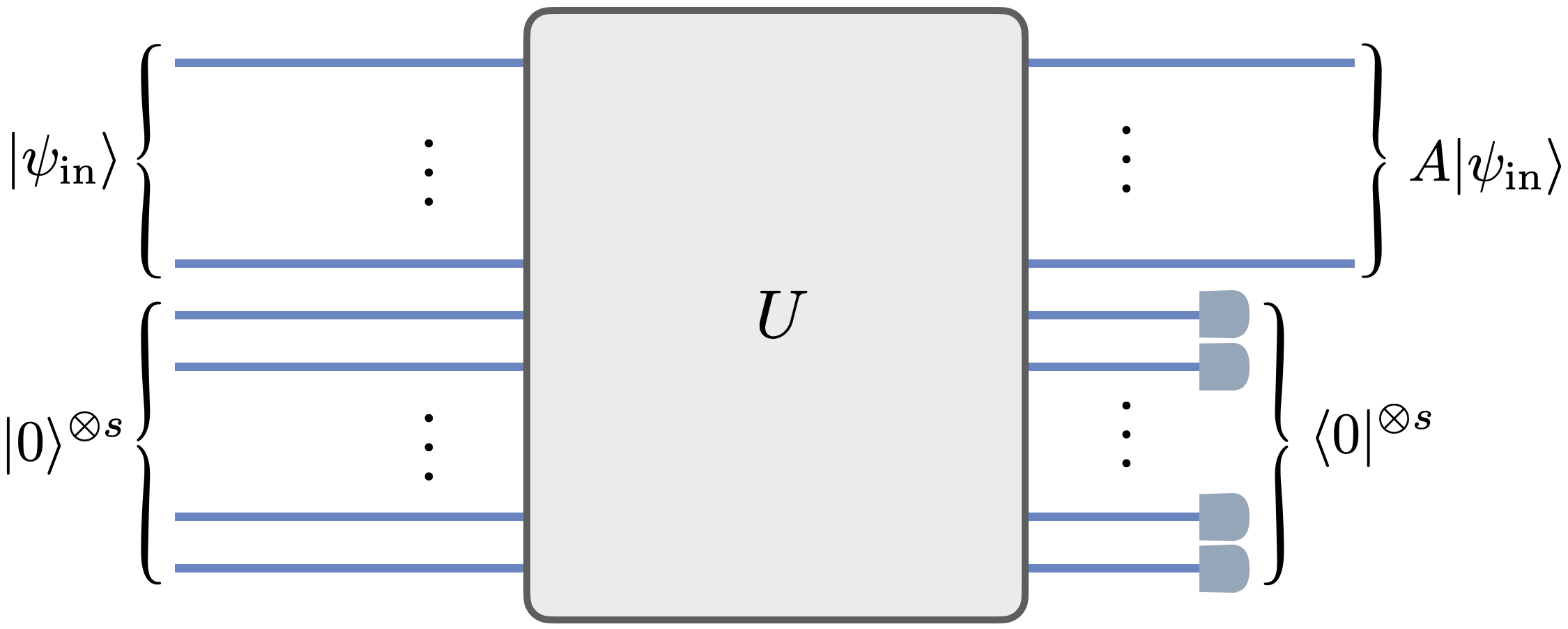}\label{fig:a}}
    
    \vspace{1em} 
    
    {\includegraphics[width=0.45\textwidth]{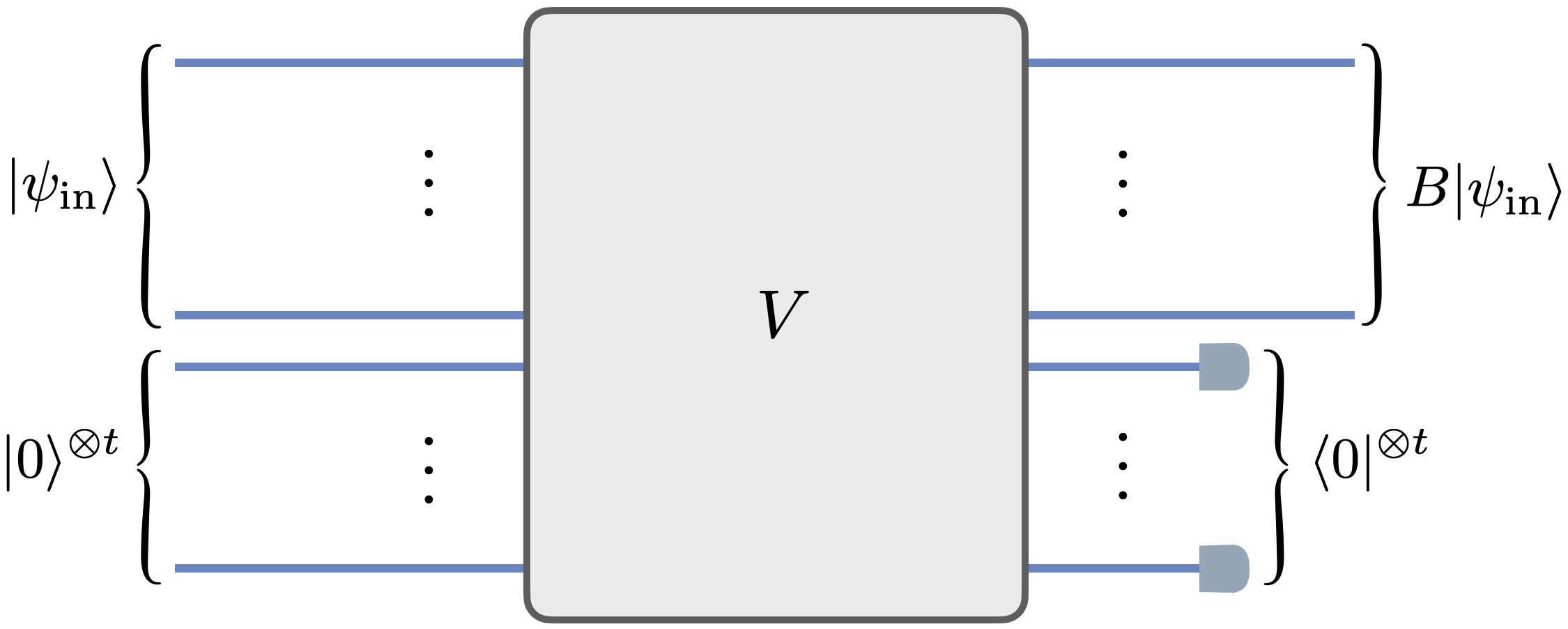}}\label{fig:a}}
    \caption{Schematic representation of the quantum circuits considered in Theorem 1. (a) Circuit implementing linear transformation $A \in \mathit{GL}(m, \mathbb{C})$ with unitary $U$ and $s$ vacuum auxiliary modes such that $\Vert A \Vert \leq 1$. (b) Circuit implementing linear transformation $B \in \mathit{GL}(m, \mathbb{C})$ with unitary $V$ and $t$ vacuum auxiliary modes such that $\Vert B \Vert \leq 1$. $\vert \psi_{\mathrm{in}} \rangle $ represents the fixed input state vector in both circuits. The vacuum auxiliary modes ($\vert 0 \rangle^{\otimes s}$ and $\vert 0 \rangle^{\otimes t}$, where $s, t \geq 0$) are heralded onto vacuum in the measurement step, resulting in the application of transformations $A$ and $B$ to $\vert \psi_{\mathrm{in}} \rangle $ with non-zero heralding probabilities $p_A$ and $p_B$, respectively. When $s = 0$ or $t = 0$, the corresponding circuit reduces to a standard linear optical evolution, such that $p_A = 1$ or $p_B = 1$, respectively. 
    }
    \label{fig:vertical}
\end{figure}

To address this, we introduce a new bound on the TVD between the normalised output distributions induced by any two invertible vacuum-heralded linear transformations. The requirement of invertibility is crucial throughout this paper, as we assume—unless stated otherwise—that, like the target unitary \( U \), the vacuum-heralded transformations in consideration, including \( U_{\text{avg}} \), are invertible and thus belong to the general linear group \( \mathit{GL}(m, \mathbb{C}) \). This assumption is essential for leveraging the Schur-Weyl duality in our methods (see Section~\ref{proof}), which applies to \( \mathit{GL}(m, \mathbb{C}) \) but not to arbitrary matrices in \( \mathbb{C}^{m \times m} \). This framework, illustrated in Fig.~\ref{fig:vertical}, forms the foundation for Theorem~\ref{theorem1}.

\begin{theorem}[Closeness of distributions]
Consider two copies of an input state consisting of $n$ photons in $m$ modes, sent into unitary interferometers $U$ and $V$, along with $s \geq 0$ and $t \geq 0$ vacuum auxiliary modes. Let heralding on vacuum on these auxiliary modes implement general linear transformations $A, B \in \mathit{GL}(m,\mathbb{C})$, respectively, ensuring that $\Vert A \Vert \leq 1$ and $\Vert B \Vert \leq 1$. Assuming that the success probabilities $p_A$ and $p_B$ of implementing $A$ and $B$ are positive (i.e., $p_A, p_B > 0$), it follows that $\max(\Vert A p_A^{-1/2n} \Vert, \Vert B p_B^{-1/2n} \Vert) \leq k$ for some $k \geq 1$. Under these conditions, the total variation distance between the $n$-photon $m$-mode boson sampling distributions $\mathcal{D}_A$ and $\mathcal{D}_B$, induced by $A$ and $B$ respectively, is bounded by
\begin{equation}
    \Vert \mathcal{D}_A - \mathcal{D}_B \Vert \leq n k^{n-1} \Vert A p_A^{-1/2n} - B p_B^{-1/2n} \Vert_{\mathrm{op}}.
\end{equation}
\label{theorem1}
\end{theorem}

For the $N$-level boson sampling averaging setup in consideration, we may assume $A = U$, $B = U_{\mathrm{avg}} =\sum_{i=1}^{N} U_{i}/N$, $p_A = 1$, and $p_B = p_{\mathrm{uni}}$. We can, therefore, write an upper bound on the TVD between the distributions from the target unitary ($\mathcal{D}_{U}$) and the unitary average ($\mathcal{D}_{U_{\mathrm{avg}}}$) as
\begin{align}
    \Vert \mathcal{D}_{U} - \mathcal{D}_{U_{\mathrm{avg}}} \Vert \text{} \leq \text{} n k^{n-1} \Bigg\Vert U - \frac{ U_{\mathrm{avg}} }{p_{\mathrm{uni}}^{1/2n}} \Bigg\Vert_{\mathrm{op}}, 
\end{align}
where $k \coloneq \max(\Vert U \Vert, \Vert U_{\text{avg}} p_{\text{post}}^{-1/2n} \Vert)$. 
Therefore, in the limit of infinite averaging ($N \rightarrow \infty$), using Eq.\ \eqref{eq:finalstate} we can write
\begin{equation}
    \Vert \mathcal{D}_{U} - \mathcal{D}_{U_{\mathrm{avg}}} \Vert \leq 0,
    \label{eq:uni_avg_bound}
\end{equation} 
establishing the convergence of the output state of the unitary averaging framework to the target state.

\begin{figure*}[!htb]
    \centering
    {\includegraphics[width=0.42\textwidth]{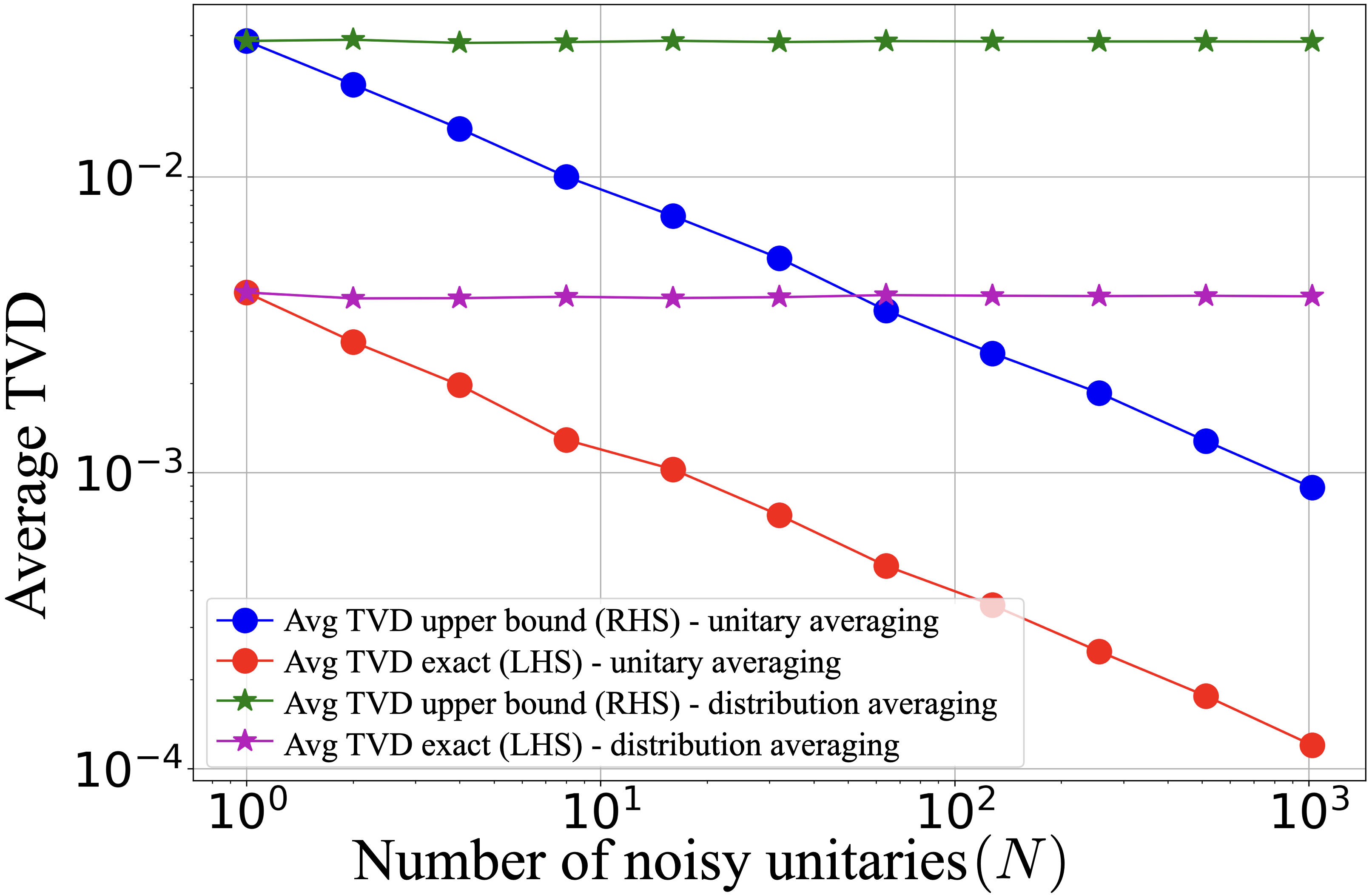}}
    \hfill
    {\includegraphics[width=0.446\textwidth]{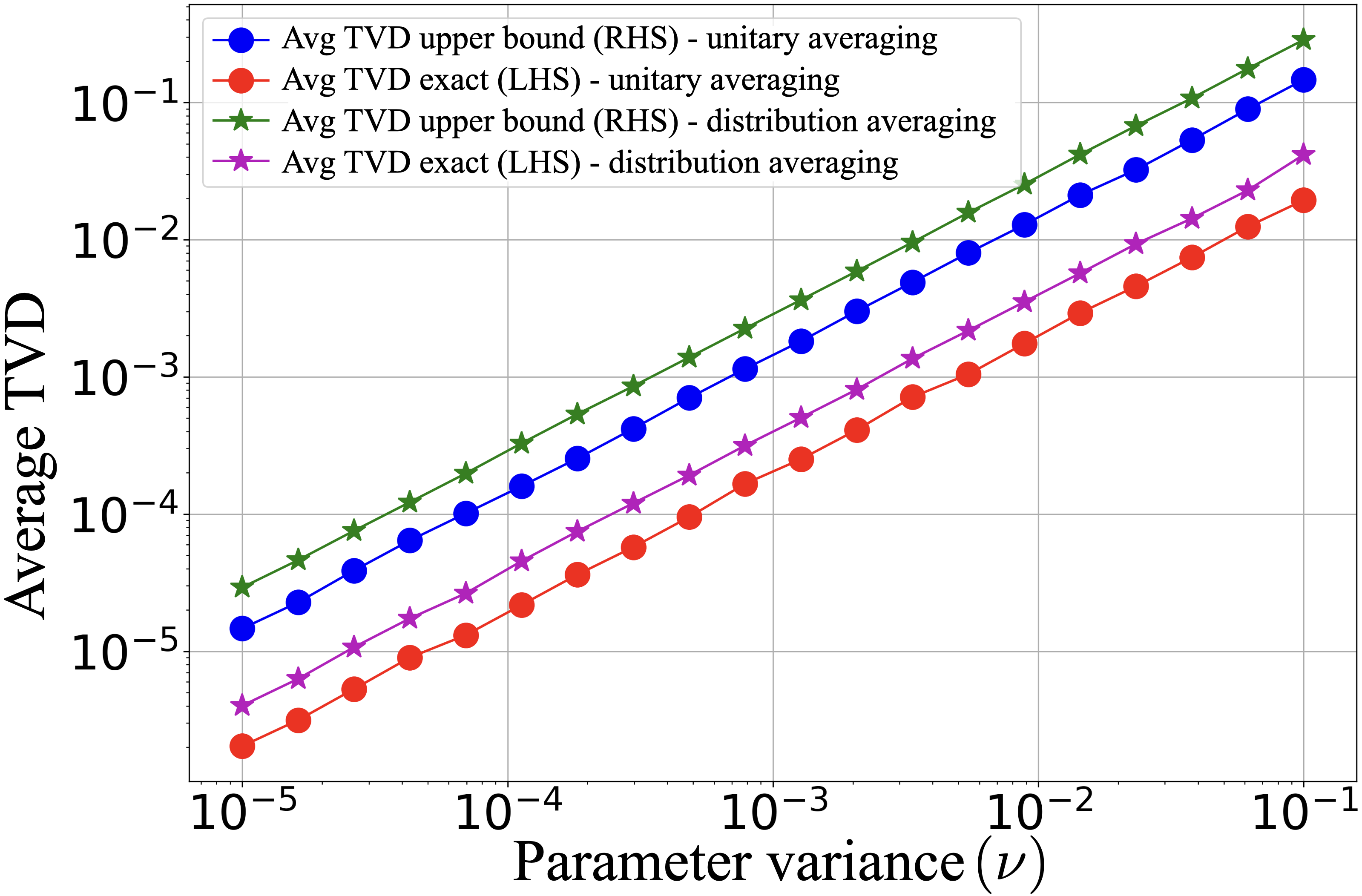}}
    \\
    {\includegraphics[width=0.42\textwidth]{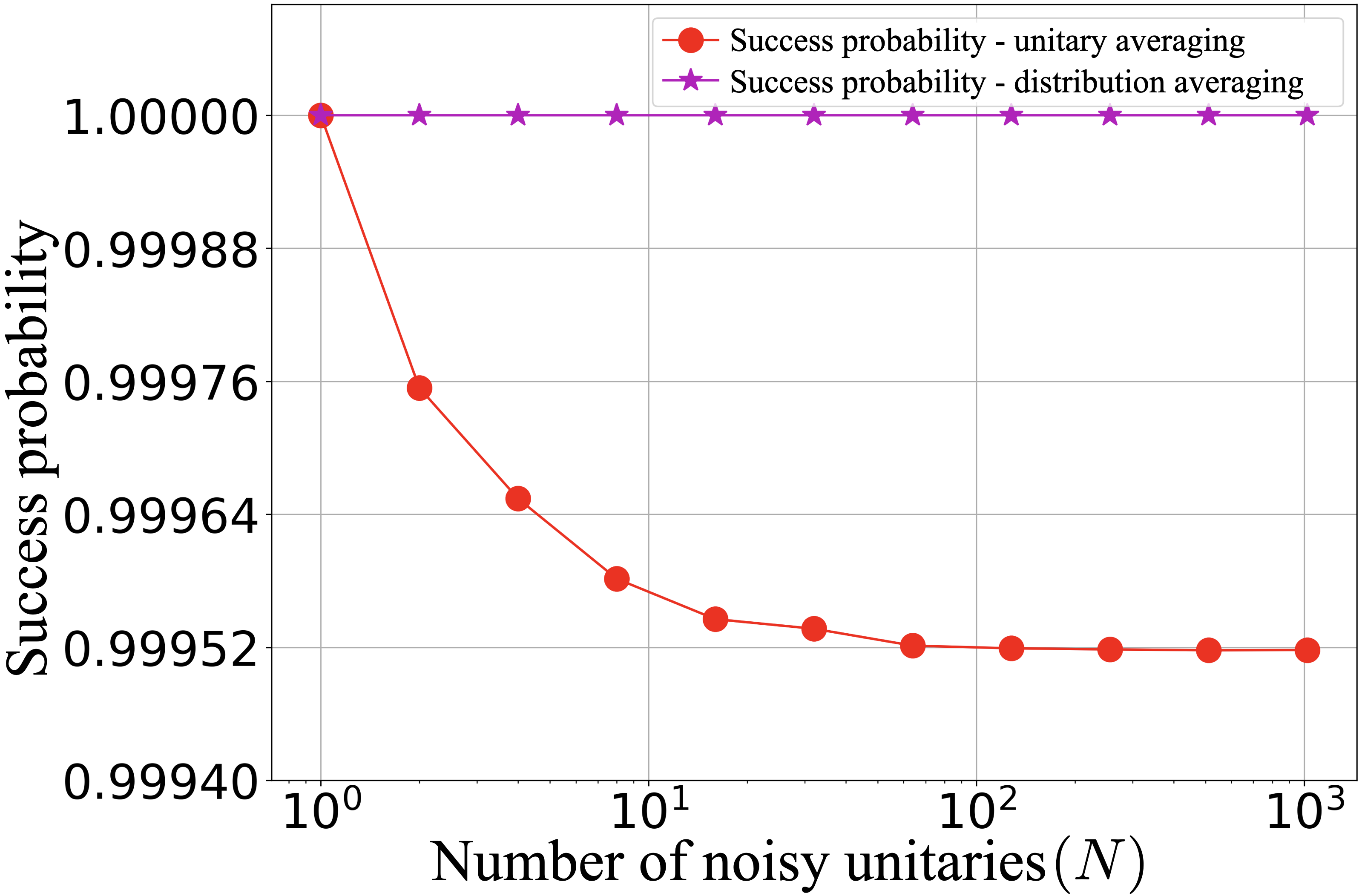}}
    \hfill
    {\includegraphics[width=0.42\textwidth]{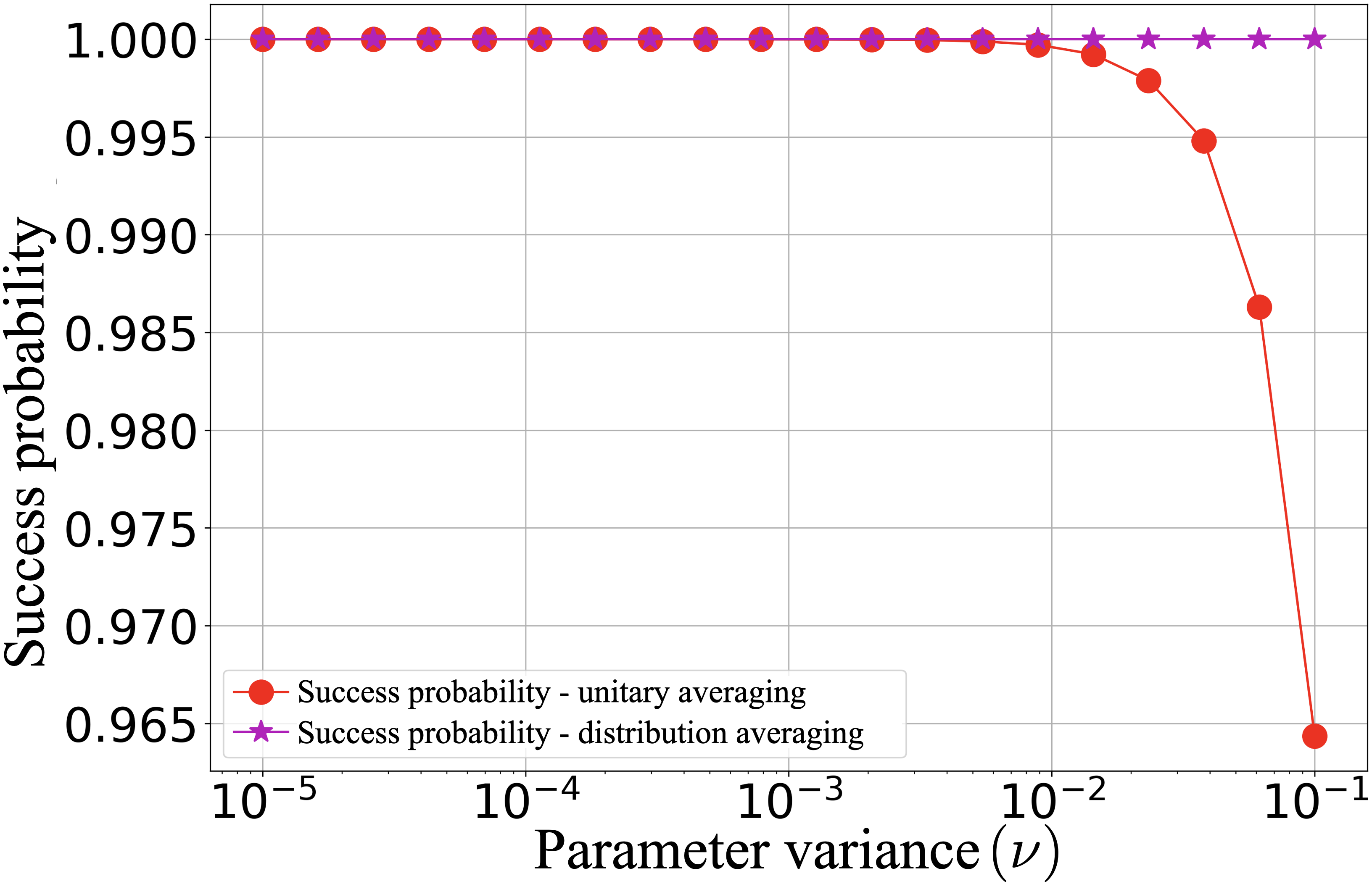}}
    \caption{Comparison of distribution averaging and unitary averaging protocols for noisy linear networks, averaged over 300 Monte Carlo runs. The setup consists of two single-photon inputs to a 2-mode unitary. (a) Average \emph{total variation distance}  (TVD) vs. the number of redundant noisy unitaries ($N$) at a fixed error variance ($\nu = 0.01$). (b) Average TVD vs. parameter variance ($\nu$) for a fixed number of noisy unitaries ($N = 4$). (c) Success probability $p_{\mathrm{post}}$ of unitary averaging as $N$ increases. The success probability of distribution averaging remains constant at 1, as it is deterministic. (d) Success probability $p_{\mathrm{post}}$ of unitary averaging vs $\nu$, with distribution averaging maintaining a deterministic success probability of 1. All plots show exact TVD values and upper bounds for both methods. Lines with the circle ($o$) markers represent unitary averaging results, while lines with star (*) markers indicate distribution averaging outcomes. As $N$ grows, the distribution from the unitary averaging protocol converges to the target distribution, which comes at the cost of a reduced success probability.}
    \label{fig:sidebyside}
\end{figure*}

We compare the performances of the distribution and unitary averaging protocols in Fig.~\ref{fig:sidebyside}, showing that although the lower bound on the success probability becomes exponentially small with the number of photons and the interferometer depth, the actual success probability can scale significantly better. In this particular case, with a small photon number, shallow depth, and low noise variance, the true success probability is at least an order of magnitude higher than the lower bound.

\section{Proof of result}
\label{proof}

This section presents the proof of Theorem~\ref{theorem1}, divided into two parts. In Section~\ref{5A}, we establish an upper bound on the distance between homomorphisms of general linear matrices as a function of the distance between the matrices themselves. In Section~\ref{5B}, we relate this homomorphism distance to the distance between the associated boson sampling distributions. 

\subsection{Convergence of homomorphisms}
\label{5A}

In this subsection, we aim to prove the bound
\begin{equation}
\Vert \phi(A) - \phi(B) \Vert \leq n k^{n-1} \Vert A - B \Vert,
\end{equation}
where \(A, B \in \mathit{GL(m, \mathbb{C})}\), \(k = \max\{\Vert A \Vert, \Vert B \Vert\}\), and $\phi$ is the homomorphism from a unitary acting on one boson to that acting on $n$ identical bosons \cite{AaransonArkhipov, Arkhipov_2015, Lund2023}. 
Using the Schur-Weyl duality for the vector space \(\mathcal{V}\), the decomposition is given by
\begin{equation}
\mathcal{V}^{\otimes n} \simeq \mathit{S^n\mathcal{V}} \oplus \mathit{\Lambda^n\mathcal{V}} \oplus \dots,
\end{equation}
where \(\mathit{S^n\mathcal{V}}\) and \(\mathit{\Lambda^n\mathcal{V}}\) denote the completely symmetric and anti-symmetric subspaces, respectively. 
Sub-representations with mixed symmetries are omitted for brevity. For any matrix \(M \in \mathit{GL(m, \mathbb{C})}\) acting on the vector space $\mathcal{V}$, we can write the block decomposition
\begin{equation}
    P M^{\otimes n} P^\dagger = \phi(M) \oplus \mu(M) \oplus \dots, \label{eq:decomposition}
\end{equation}
where \(P\) is a suitable unitary change-of-basis matrix, and \(\phi(M)\) and \(\mu(M)\) correspond to the 
symmetric and anti-symmetric sub-representations of \(M\).

We will also later use the fact that for any block diagonal matrix \(Q = \bigoplus_{i=1}^L Q_i\), the operator norm satisfies
\begin{equation}
    \Vert Q \Vert = \max_{1 \leq i \leq L} \Vert Q_i \Vert \implies \Vert Q_i \Vert \leq \Vert Q \Vert, \quad \forall 1 \leq i \leq L. \label{eq:inequality}
\end{equation}
Employing the unitary invariance of the operator norm, for any two matrices \(A\) and \(B\), we have
\begin{equation}
\Vert P A^{\otimes n} P^\dagger - P B^{\otimes n} P^\dagger \Vert = \Vert A^{\otimes n} - B^{\otimes n} \Vert,
\end{equation}
which can be rewritten using 
Eq.\ \eqref{eq:decomposition}, as
\begin{equation}
\Vert A^{\otimes n} - B^{\otimes n} \Vert = \Vert (\phi(A) - \phi(B)) \oplus (\mu(A) - \mu(B)) \oplus \dots \Vert.
\end{equation}
Using Eq.\ \eqref{eq:inequality}, this simplifies to
\begin{equation}
\Vert \phi(A) - \phi(B) \Vert \leq \Vert A^{\otimes n} - B^{\otimes n} \Vert.
\end{equation}
To bound the norm of \(A^{\otimes n} - B^{\otimes n}\), we expand the tensor powers as
\begin{align}
\Vert A^{\otimes n} - B^{\otimes n} \Vert &= \Vert A^{\otimes n} - A^{\otimes n-1}B + A^{\otimes n-1}B - \dots \\
\nonumber
& \quad + AB^{\otimes n-1} - B^{\otimes n} \Vert \\
\nonumber
& \leq \Vert A^{\otimes n-1} \Vert \Vert A - B \Vert + \Vert A^{\otimes n-2}B \Vert \Vert A - B \Vert \\
\nonumber
& \quad + \dots + \Vert B^{\otimes n-1} \Vert \Vert A - B 
\Vert.
\nonumber
\end{align}
Using the property \(\Vert M_1 \otimes M_2 \Vert = \Vert M_1 \Vert \Vert M_2 \Vert\) for any two matrices \(M_1\) and \(M_2\), and choosing \(\Vert A \Vert \leq \Vert B \Vert \leq k\), we can simplify this to
\begin{equation}
\Vert A^{\otimes n-1} \Vert, \Vert A^{\otimes n-2}B \Vert, \dots, \Vert B^{\otimes n-1} \Vert \leq k^{n-1},
\end{equation}
which leads to
\begin{equation}
\Vert A^{\otimes n} - B^{\otimes n} \Vert \leq n k^{n-1} \Vert A - B \Vert,
\end{equation}
which in turn proves the inequality 
\begin{equation}
\Vert \phi(A) - \phi(B) \Vert \leq n k^{n-1} \Vert A - B \Vert.
\end{equation}

\subsection{Convergence of distributions}
\label{5B}

Let $\vert \psi_{{\mathrm{out}},A} \rangle $ be the normalised output state vector obtained by applying the linear transformation $A$ on any given normalised input state vector $\vert \psi_{\mathrm{in}} \rangle$. Let $p_A$ denote the success probability of this transformation such that $p_A = \langle \psi_{\mathrm{in}} \vert {\phi(A)}^\dagger \phi(A) \vert \psi_{\mathrm{in}} \rangle$. Similarly, for B. Then, 
\begin{align*}
    \Big\Vert \vert\psi_{{\mathrm{out}},A}\rangle - \vert\psi_{{\mathrm{out}},B}\rangle \Big\Vert &= \Big\Vert \frac{\phi(A)}{\sqrt{p_A}}\vert\psi_{\mathrm{in}}\rangle - \frac{\phi(B)}{\sqrt{p_B}}\vert\psi_{\mathrm{in}}\rangle \Big\Vert, \\
    &= \Big\Vert \frac{\phi(A)}{\sqrt{p_A}} - \frac{\phi(B)}{\sqrt{p_B}}\Big\Vert. \numberthis
\end{align*}
Since $\phi(M)$ for any general linear matrix $M$ can be expressed as a matrix whose entries are symmetric polynomials in the entries of $M$, we can write
\begin{align}
    \frac{\phi(M)}{\sqrt{p_M}} = \phi\Big(\frac{M}{p_M^{1/2n}}\Big).  
\end{align}
Redefining the matrices, we can write $A^{'} = {A}/{p_A^{1/2n}}$ and $B^{'}={B}/{p_B^{1/2n}}$. 
Using the results from Sec.\ \ref{5A}, we get
\begin{align*}
    \Big\Vert \vert\psi_{{\mathrm{out}},A}\rangle - \vert\psi_{{\mathrm{out}},B}\rangle \Big\Vert &= \Big\Vert \phi\Big(\frac{A}{p_A^{1/2n}}\Big) - \phi\Big(\frac{B}{p_B^{1/2n}}\Big) \Big\Vert \\
    &\leq n k^{n-1}\Big\Vert \frac{A}{p_A^{1/2n}} - \frac{B}{p_B^{1/2n}} \Big\Vert. \numberthis 
\end{align*}
Let $\mathcal{D}_A$ and $\mathcal{D}_B$ be the distributions generated by state vectors $\vert\psi_{{\mathrm{out}},A}\rangle$ and $\vert\psi_{{\mathrm{out}},B}\rangle$, respectively. Since
\begin{align}
\Vert \mathcal{D}_A - \mathcal{D}_B \Vert \leq \Big\Vert \vert\psi_{{\mathrm{out}},A}\rangle - \vert\psi_{{\mathrm{out}},B}\rangle \Big\Vert,
\end{align}
we finally have
\begin{equation}
    \Vert \mathcal{D}_A - \mathcal{D}_B \Vert \leq n k^{n-1} \Bigg\Vert \frac{A}{p_A^{1/2n}} - \frac{B}{p_B^{1/2n}} \Bigg\Vert_{\mathrm{op}}.
    \label{eq:final_bound}
\end{equation} 

\section{Discussion}
\label{discussion}

Noise and loss remain fundamental challenges in demonstrating quantum advantage in boson sampling experiments. Loss impacts the scalability of these experiments by reducing the Hilbert space of the output sample, thereby simplifying simulation by classical algorithms. However, loss is more straightforward to detect during post-processing. In contrast, noise poses a more significant challenge, as it is harder to identify without a detailed characterisation of the single-photon sources, interferometers, and detectors used in the setup. This raises critical questions about the implementation's fidelity for the desired Haar-random unitary, indistinguishable single photons, and ideal detection. Previous works have explored the effects of photon distinguishability and loss on boson sampling experiments, identifying the conditions under which quantum advantage can actually be achieved.

Among the various noise sources affecting boson samplers, Arkhipov \cite{Arkhipov_2015} quantified the impact of unitary errors on the resulting boson sampling distribution. Building on this, Theorem \ref{theorem1} extends their result to vacuum-heralded boson samplers, providing a quantitative understanding of how unitary errors influence their output distributions. This generalisation points toward a broader framework for analysing state generation methods in linear optics—originally introduced by the KLM protocol~\cite{knill2001klm}.

As an application of Theorem \ref{theorem1}, this study addresses the critical issue of mitigating unitary noise in boson samplers by developing and characterising the unitary averaging protocol. We focus on the experimentally relevant model of stochastic unitary noise and prove that the mitigated distribution converges to the ideal boson sampling distribution as the encoding in the unitary averaging protocol increases. This result is significant as it demonstrates that the desired Haar-random unitary can be effectively implemented at the expense of a manageable sampling overhead.

Our proposed mitigation approach is particularly relevant since recent advances in the field have enabled linear optical interferometers to achieve 
high fidelity \cite{alexander_manufacturable_2025}. While the lower bound on the success probability of the mitigation scheme decreases exponentially with the number of photons and the depth of the boson sampler, it is reasonable to assume that any practical interferometers will exhibit minor parameter variances, leaving the associated sampling overhead feasible. This work, therefore, presents a promising strategy to advance efforts in demonstrating quantum advantage with contemporary boson samplers. Given the exponential scaling of the lower bound on the success probability, this scheme is best suited for intermediate-scale experiments. 

However, as highlighted in~\cite{ryan2024, PhysRevApplied.23.044003}, large-scale photonic quantum computing experiments may also benefit from integrating near-term mitigation strategies—targeting unitary errors, photon distinguishability, and related imperfections—with conventional error correction codes and fault-tolerance schemes. Such integration could reduce the resource overhead of quantum computations~\cite{PhysRevApplied.23.044003} or improve the error thresholds of the chosen error correction codes~\cite{ryan2024}. Since the unitary averaging protocol works by converting logical errors into heralded loss, the use of parity encoding~\cite{PhysRevLett.95.100501,PhysRevA.77.012310} has been suggested by~\cite{ryan2024} to protect against this photon loss.  

From an experimental implementation perspective, it is noteworthy that the effects of encoding and decoding errors on unitary averaging can be suppressed to first order~\cite{ryan2024}. This observation justifies focusing solely on stochastic errors in the noisy unitaries, while safely neglecting the impact of encoding and decoding imperfections. Furthermore, the encoding and decoding unitaries can be implemented with a depth that scales logarithmically with the amount of averaging, $N$, remaining well within reach for intermediate-scale experimental demonstrations, as originally intended. 

\section{Further Applications}
\label{furtherConsiderations}

This section explores some potential implications and applications of our work. We begin by discussing how our theorem can offer an exponential advantage in estimating the distance between two invertible vacuum-heralded transformations when the corresponding bound is tight. We then demonstrate how the unitary averaging framework emerges as a special case of the \emph{linear combination of unitaries} (LCU) implementation in linear optics and present a general scheme for arbitrary LCU implementation. Furthermore, we highlight how unitary averaging can also act as a tool for benchmarking the repeatability of experimental implementations of given unitaries in linear optics, providing a new method to validate the consistency of an experiment. Lastly, we analyse the broad hierarchy of transformations enabled by linear optics when augmented with vacuum heralding and single-photon transformations.

\subsection{Estimating the computational effort}

We consider the computational complexity of computing the distance between the distributions induced by two invertible vacuum-heralded linear optical networks. Assuming that a total of $n$ photons evolve through known $m$-mode heralded linear transformations $A$ and $B$, computing each distribution $\mathcal{D}_A$ and $\mathcal{D}_B$ takes $O\left({{m+n-1}\choose{n}} n2^n\right)$ time. Here, $O(n2^n)$ is the complexity of computing a single matrix permanent using Ryser's algorithm \cite{ryser} and ${{m+n-1}\choose{n}}$ is the number of possible configurations. Therefore, the time complexity to compute the exact distance between the distributions induced by $A$ and $B$ is $O({{m+n-1}\choose{n}} n2^n)$. In comparison, to find the upper bound as defined in Theorem \ref{theorem1}, one can compute probabilities $ p_A$ and $p_B$ each in only $O(\log(n)n^3 2^n)$, using the recent results on coarse-grained boson sampling distributions \cite{seron2022efficientvalidationbosonsampling}. Considering the polynomial cost of computing the operator norm, the upper bound on the distance between the distributions induced by any two vacuum-heralded linear optical networks 
can be computed 
with exponentially less computational effort than 
the exact distance. 

\subsection{Linear combination of unitaries in linear optics}
It is interesting to note that since any $m \times m$ complex matrix can be written as a linear combination of a maximum of four unitaries \cite{PhysRevLett.127.270503}, the UA protocol also provides a method to implement any linear transformation in optics. This can be done by replacing the \emph{discrete Fourier transform} (DFT) matrices in the UA framework with appropriate encoding and decoding matrices $E$ and $D$, so the unitaries can be combined with the required coefficients. Specifically, assuming unitary averaging with $N$ redundant copies of $m$ mode unitaries $U_i$, where $1 \leq i \leq N$, we can write the evolution of the creation operators through the encoding matrix as 
\begin{equation}
       a_{j,r} \mapsto \sum\limits_{k=0}^{N-1} E_{r,k}a_{j,k}, 
\end{equation}
where, $1\leq j \leq m$ and $0 \leq r \leq N-1$, such that in the notation used, $a_{j,0}$ are the original input modes and $a_{j,i}$, such that $i \in \{1,2,\dots, N-1\}$, are the auxiliary modes. Further evolution through the redundant unitaries and the decoder matrices induces the following transformation
\begin{equation}
   a_{j,r} \mapsto \sum\limits_{l=0}^{m-1} \sum\limits_{k,k'=0}^{N-1} (U_{k'})_{l,j} E_{r,k'} D_{k,k'}a_{l,r}.
\end{equation}
Upon heralding on the vacuum in the auxiliary modes, the evolution of the original modes can be written by choosing $r,k=0$, giving
\begin{equation}
       a_{j,0} \mapsto \sum\limits_{l=0}^{m-1} \sum\limits_{k'=0}^{N-1} (U_{k'})_{l,j} E_{0,k'} D_{0,k'}a_{l,0}. 
\end{equation}
Finally, the complete transformation through the unitary averaging network can be written as 
\begin{equation}
       a_{j,0} \mapsto \sum\limits_{l=0}^{m-1} \left( M_N\right)_{l,j}a_{l,0}, 
\end{equation}
such that $M_N$ is defined as 
\begin{equation}
    M_N = \sum\limits_{k'=0}^{N-1} U_{k'} E_{0,k'} D_{0,k'} = \sum\limits_{k'=0}^{N-1} \alpha_{k'} U_{k'}, 
\end{equation}
where the encoder $E$ and decoder $D$ can be set such that $\alpha_{k'} = E_{0,k'} 
D_{0,k'}$ such that $\alpha_{k'}$ are the required coefficients for any given linear combination of unitaries. 

This result is essential for implementing the \emph{linear combination of unitaries} (LCU) \cite{10.5555/2481569.2481570} 
and \emph{quantum singular value transform} (QSVT) \cite{10.1145/3313276.3316366} framework in practice. Unitary averaging, in essence, provides a general framework for the photonic implementation of quantum algorithms utilising these techniques using passive linear optical elements and vacuum heralding only.

\begin{figure}[htbp]
    \centering
  \includegraphics[width=0.35\textwidth]{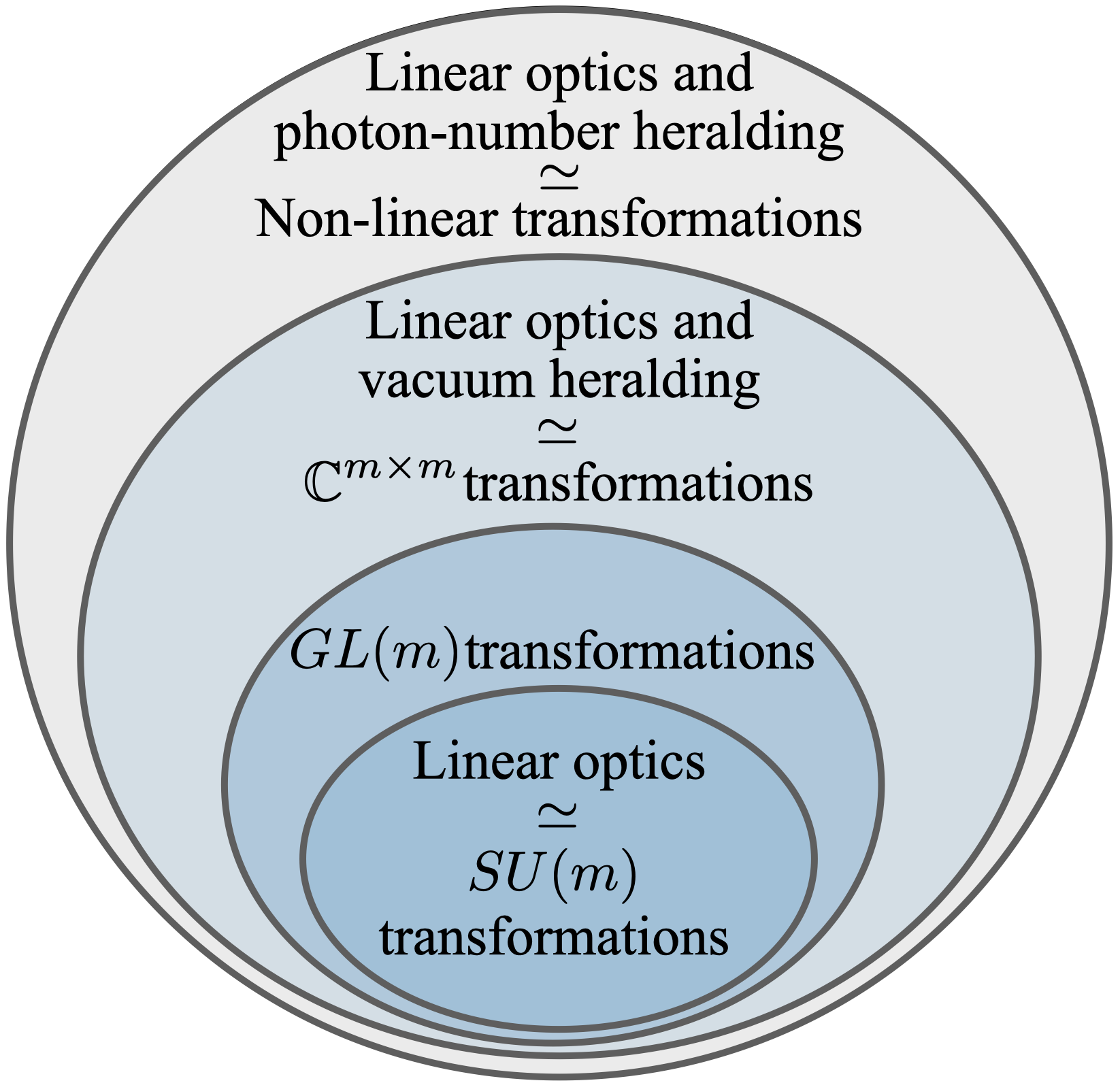}
    \caption{This figure illustrates the hierarchy of quantum state preparation capabilities in linear optics. At its core, basic linear optics allows for $SU(m)$ transformations in an $m$-mode interferometer. The next level, achieved by incorporating vacuum heralding on ancillary modes, expands this to $\mathbb{C}^{m \times m}$ transformations, significantly broadening the set of achievable states. The outermost level, which includes non-zero photon-number heralding, enables universal quantum transformations as demonstrated by the KLM (\emph{Knill-Laflamme-Milburn}) theorem, and matrices can not represent its action in general because of the involved non-linear transformations. 
    We also highlight the regime of $\mathit{GL}(m)$ transformations for which Theorem \ref{theorem1} is valid.} 
    \label{fig:hierarchy}
\end{figure}

\subsection{Benchmarking fabrication repeatability}
Achieving precise control over the fabrication of interferometers in chips or wafers to implement desired unitary transformations remains a significant challenge, raising concerns about the repeatability of these processes. In general, variations in interferometers built using the same fabrication process can arise from systematic deviations, process drift, or random variations \cite{SAGHAEI2019100733}. Given this inherent variability, developing methods for verifying whether a set of physical interferometers performs the same linear transformation is crucial.

The unitary averaging protocol can be effectively adapted to address this challenge. Consider $ N$ arbitrary interferometers, $ U_1, U_2, \dots, U_N$, fabricated using the same process. These interferometers can be incorporated into the unitary averaging protocol with single photons input in the setup. A sampling task is then performed, where all output modes are measured, and the occurrence of non-zero photon counts in the ancilla modes is recorded. The presence of such counts acts as a witness to a discrepancy in one or more chips. 

When all chips implement the same interferometer, the unitary averaging protocol ensures that all input photons remain confined to the original modes, resulting in vacuum states in the ancilla modes. This outcome is a direct consequence of the protocol: upon vacuum heralding in the ancilla modes, the transformation $\sum_i U_i / N$ is applied. If $U_1 = U_2 = \dots = U_N$, this transformation is unitary, preventing any input photons from routing to the ancilla modes.
Thus, the probability of non-zero photon counts in the ancilla modes is a quantitative indicator of fabrication non-repeatability. This method provides an alternative to the direct characterisation of all the interferometers to benchmark their closeness \cite{Rahimi-Keshari:13,Laing:2012,Hoch:2023,Fyrillas:2024}.

\subsection{Hierarchy of transformations in linear optics}

Linear optics allows only a limited set of transformations in the possible state transformation space. This limitation becomes particularly critical when considering the requirements for universal quantum computation or implementing specific quantum communication protocols, which often necessitate more complex, non-linear transformations. While non-linear optical processes can, in principle, provide the required transformations, their practical implementation remains challenging due to weak non-linearities in conventional optical materials and the difficulty of maintaining quantum coherence.

A breakthrough in addressing this challenge came with the insight that measurements can, in effect, generate the required non-linearities in linear optical schemes
\cite{RevModPhys.79.135,PhysRevLett.95.010501,PhysRevLett.96.020501,knill2001klm}, an insight that can be traced back to the seminal work of Knill, Laflamme, and Milburn \cite{knill2001klm}. Their 
theorem has demonstrated that it is -- at least in principle 
-- possible to simulate non-linear transformations using linear optics, supplemented by photon detection and feed-forward operations. This process of probabilistically generating the target state based on a particular photon detection pattern in the auxiliary modes, known as photon heralding, effectively expands the realm of accessible photonic states beyond what is achievable with linear optics alone. Later schemes have made this core idea that is at the heart of linear optical quantum computing substantially more resource-efficient \cite{RevModPhys.79.135,PhysRevLett.95.010501,PhysRevLett.96.020501,Fusion, PhysRevA.74.042343}.

In this work, we derive bounds on distances between linear optical systems with access to vacuum heralding. Although this setup does not provide access to the complete space of multi-photon multi-mode states, it is more powerful than linear optics alone, as depicted in Fig.~\ref{fig:hierarchy}. Although this hierarchy has been known before \cite{Park2023}, the exact transformation belonging to each class has not been highlighted. We quantify 
such transformations in Fig.~\ref{fig:hierarchy} 
and provide rigorous limits on the ability of vacuum heralding in linear optics.\\ 

\section{Conclusions and outlook} \label{conclusion}

In this work, we extend the concept of error mitigation to sampling tasks through a new variant of the unitary averaging protocol. By leveraging multiple independent and identical stochastic boson samplers, we demonstrate the ability to generate a distribution that converges to 
the ideal boson sampling distribution as the number of samplers increases, with a success probability that is lower-bounded by an exponential function of the number of photons and the depth of the boson sampler. Given this demanding scaling, reminiscent of the situation in standard quantum error mitigation, we identify regimes where unitary averaging might offer an advantage over contemporary classical sampling algorithms. While we have confined the discussion to stochastic errors, again, it is plausible to expect the scheme to also mitigate against a larger class of errors. The convergence of the unitary averaging distribution to the ideal distribution is supported by proving an upper bound on the distance between output distributions induced by two invertible vacuum-heralded linear interferometers derived using ideas from representation theory, specifically the Schur-Weyl duality. This upper bound represents a step toward quantifying the role of heralding in linear optics—a problem central to photonic quantum computation, given the fundamental role of heralding in photonic state generation as emphasised by the KLM protocol.

We further explore the enhanced capabilities introduced by vacuum heralding in linear optics. In particular, we quantify the broader class of transformations achievable with vacuum heralding compared to standard linear optics alone. Building on this, we extend the unitary averaging protocol to realize the \emph{linear combinations of unitaries} (LCU) framework using only passive optical elements and vacuum heralding. This demonstrates the power of probabilistic techniques in overcoming inherent limitations in photonic quantum systems, opening the door to more resilient quantum protocols. More broadly, we aim to advance the development of meaningful quantum error mitigation strategies for sampling problems and other applications that extend beyond mere expectation value estimation.

\section*{Acknowledgements}

DS appreciates help from Marco Tomamichel in the proof of the main theorem.  This work has been supported by the BMBF (PhoQuant, QCIP/1), QuantERA (HQCC), BMWK (EniQmA), the Quantum Flagship (PasQuans2, Millenion), the Munich Quantum Valley, 
Berlin Quantum, European Research Council (DebuQC), and the Australian Research Council through the
Centre of Excellence for Quantum Computation and Communication Technology (Project No. CE170100012). 

\bibliography{references} 

\begin{thebibliography}{53}%
\makeatletter
\providecommand \@ifxundefined [1]{%
 \@ifx{#1\undefined}
}%
\providecommand \@ifnum [1]{%
 \ifnum #1\expandafter \@firstoftwo
 \else \expandafter \@secondoftwo
 \fi
}%
\providecommand \@ifx [1]{%
 \ifx #1\expandafter \@firstoftwo
 \else \expandafter \@secondoftwo
 \fi
}%
\providecommand \natexlab [1]{#1}%
\providecommand \enquote  [1]{``#1''}%
\providecommand \bibnamefont  [1]{#1}%
\providecommand \bibfnamefont [1]{#1}%
\providecommand \citenamefont [1]{#1}%
\providecommand \href@noop [0]{\@secondoftwo}%
\providecommand \href [0]{\begingroup \@sanitize@url \@href}%
\providecommand \@href[1]{\@@startlink{#1}\@@href}%
\providecommand \@@href[1]{\endgroup#1\@@endlink}%
\providecommand \@sanitize@url [0]{\catcode `\\12\catcode `\$12\catcode `\&12\catcode `\#12\catcode `\^12\catcode `\_12\catcode `\%12\relax}%
\providecommand \@@startlink[1]{}%
\providecommand \@@endlink[0]{}%
\providecommand \url  [0]{\begingroup\@sanitize@url \@url }%
\providecommand \@url [1]{\endgroup\@href {#1}{\urlprefix }}%
\providecommand \urlprefix  [0]{URL }%
\providecommand \Eprint [0]{\href }%
\providecommand \doibase [0]{https://doi.org/}%
\providecommand \selectlanguage [0]{\@gobble}%
\providecommand \bibinfo  [0]{\@secondoftwo}%
\providecommand \bibfield  [0]{\@secondoftwo}%
\providecommand \translation [1]{[#1]}%
\providecommand \BibitemOpen [0]{}%
\providecommand \bibitemStop [0]{}%
\providecommand \bibitemNoStop [0]{.\EOS\space}%
\providecommand \EOS [0]{\spacefactor3000\relax}%
\providecommand \BibitemShut  [1]{\csname bibitem#1\endcsname}%
\let\auto@bib@innerbib\@empty
\bibitem [{\citenamefont {Flamini}\ \emph {et~al.}(2018)\citenamefont {Flamini}, \citenamefont {Spagnolo},\ and\ \citenamefont {Sciarrino}}]{Photonic}%
  \BibitemOpen
  \bibfield  {author} {\bibinfo {author} {\bibfnamefont {F.}~\bibnamefont {Flamini}}, \bibinfo {author} {\bibfnamefont {N.}~\bibnamefont {Spagnolo}},\ and\ \bibinfo {author} {\bibfnamefont {F.}~\bibnamefont {Sciarrino}},\ }\bibfield  {title} {\bibinfo {title} {Photonic quantum information processing: a review},\ }\href {https://doi.org/10.1088/1361-6633/aad5b2} {\bibfield  {journal} {\bibinfo  {journal} {Rep. Prog. Phys.}\ }\textbf {\bibinfo {volume} {82}},\ \bibinfo {pages} {016001} (\bibinfo {year} {2018})}\BibitemShut {NoStop}%
\bibitem [{\citenamefont {Kok}\ \emph {et~al.}(2007)\citenamefont {Kok}, \citenamefont {Munro}, \citenamefont {Nemoto}, \citenamefont {Ralph}, \citenamefont {Dowling},\ and\ \citenamefont {Milburn}}]{RevModPhys.79.135}%
  \BibitemOpen
  \bibfield  {author} {\bibinfo {author} {\bibfnamefont {P.}~\bibnamefont {Kok}}, \bibinfo {author} {\bibfnamefont {W.~J.}\ \bibnamefont {Munro}}, \bibinfo {author} {\bibfnamefont {K.}~\bibnamefont {Nemoto}}, \bibinfo {author} {\bibfnamefont {T.~C.}\ \bibnamefont {Ralph}}, \bibinfo {author} {\bibfnamefont {J.~P.}\ \bibnamefont {Dowling}},\ and\ \bibinfo {author} {\bibfnamefont {G.~J.}\ \bibnamefont {Milburn}},\ }\bibfield  {title} {\bibinfo {title} {Linear optical quantum computing with photonic qubits},\ }\href {https://doi.org/10.1103/RevModPhys.79.135} {\bibfield  {journal} {\bibinfo  {journal} {Rev. Mod. Phys.}\ }\textbf {\bibinfo {volume} {79}},\ \bibinfo {pages} {135} (\bibinfo {year} {2007})}\BibitemShut {NoStop}%
\bibitem [{\citenamefont {Wang}\ \emph {et~al.}(2020)\citenamefont {Wang}, \citenamefont {Sciarrino}, \citenamefont {Laing},\ and\ \citenamefont {Thompson}}]{Photonic2}%
  \BibitemOpen
  \bibfield  {author} {\bibinfo {author} {\bibfnamefont {J.}~\bibnamefont {Wang}}, \bibinfo {author} {\bibfnamefont {F.}~\bibnamefont {Sciarrino}}, \bibinfo {author} {\bibfnamefont {A.}~\bibnamefont {Laing}},\ and\ \bibinfo {author} {\bibfnamefont {M.~G.}\ \bibnamefont {Thompson}},\ }\bibfield  {title} {\bibinfo {title} {Integrated photonic quantum technologies},\ }\href {https://doi.org/10.1038/s41566-019-0532-1} {\bibfield  {journal} {\bibinfo  {journal} {Nature Phot.}\ }\textbf {\bibinfo {volume} {14}},\ \bibinfo {pages} {273} (\bibinfo {year} {2020})}\BibitemShut {NoStop}%
\bibitem [{\citenamefont {Bartolucci}\ \emph {et~al.}(2023)\citenamefont {Bartolucci}, \citenamefont {Birchall}, \citenamefont {Bombín}, \citenamefont {Cable}, \citenamefont {Dawson}, \citenamefont {Gimeno-Segovia}, \citenamefont {Johnston}, \citenamefont {Kieling}, \citenamefont {Nickerson}, \citenamefont {Pant}, \citenamefont {Pastawski}, \citenamefont {Rudolph},\ and\ \citenamefont {Sparrow}}]{Fusion}%
  \BibitemOpen
  \bibfield  {author} {\bibinfo {author} {\bibfnamefont {S.}~\bibnamefont {Bartolucci}}, \bibinfo {author} {\bibfnamefont {P.}~\bibnamefont {Birchall}}, \bibinfo {author} {\bibfnamefont {H.}~\bibnamefont {Bombín}}, \bibinfo {author} {\bibfnamefont {H.}~\bibnamefont {Cable}}, \bibinfo {author} {\bibfnamefont {C.}~\bibnamefont {Dawson}}, \bibinfo {author} {\bibfnamefont {M.}~\bibnamefont {Gimeno-Segovia}}, \bibinfo {author} {\bibfnamefont {E.}~\bibnamefont {Johnston}}, \bibinfo {author} {\bibfnamefont {K.}~\bibnamefont {Kieling}}, \bibinfo {author} {\bibfnamefont {N.}~\bibnamefont {Nickerson}}, \bibinfo {author} {\bibfnamefont {M.}~\bibnamefont {Pant}}, \bibinfo {author} {\bibfnamefont {F.}~\bibnamefont {Pastawski}}, \bibinfo {author} {\bibfnamefont {T.}~\bibnamefont {Rudolph}},\ and\ \bibinfo {author} {\bibfnamefont {C.}~\bibnamefont {Sparrow}},\ }\bibfield  {title} {\bibinfo {title} {Fusion-based quantum computation},\ }\href {https://doi.org/10.1038/s41467-023-36493-1} {\bibfield  {journal} {\bibinfo
  {journal} {Nature Comm.}\ }\textbf {\bibinfo {volume} {14}},\ \bibinfo {pages} {912} (\bibinfo {year} {2023})}\BibitemShut {NoStop}%
\bibitem [{\citenamefont {Madsen}\ \emph {et~al.}(2022)\citenamefont {Madsen}, \citenamefont {Laudenbach}, \citenamefont {Askarani}, \citenamefont {Rortais}, \citenamefont {Vincent}, \citenamefont {Bulmer}, \citenamefont {Miatto}, \citenamefont {Neuhaus}, \citenamefont {Helt}, \citenamefont {Collins}, \citenamefont {Lita}, \citenamefont {Gerrits}, \citenamefont {Nam}, \citenamefont {Vaidya}, \citenamefont {Menotti}, \citenamefont {Dhand}, \citenamefont {Vernon}, \citenamefont {Quesada},\ and\ \citenamefont {Lavoie}}]{Xanadu}%
  \BibitemOpen
  \bibfield  {author} {\bibinfo {author} {\bibfnamefont {L.~S.}\ \bibnamefont {Madsen}}, \bibinfo {author} {\bibfnamefont {F.}~\bibnamefont {Laudenbach}}, \bibinfo {author} {\bibfnamefont {M.~F.}\ \bibnamefont {Askarani}}, \bibinfo {author} {\bibfnamefont {F.}~\bibnamefont {Rortais}}, \bibinfo {author} {\bibfnamefont {T.}~\bibnamefont {Vincent}}, \bibinfo {author} {\bibfnamefont {J.~F.~F.}\ \bibnamefont {Bulmer}}, \bibinfo {author} {\bibfnamefont {F.~M.}\ \bibnamefont {Miatto}}, \bibinfo {author} {\bibfnamefont {L.}~\bibnamefont {Neuhaus}}, \bibinfo {author} {\bibfnamefont {L.~G.}\ \bibnamefont {Helt}}, \bibinfo {author} {\bibfnamefont {M.~J.}\ \bibnamefont {Collins}}, \bibinfo {author} {\bibfnamefont {A.~E.}\ \bibnamefont {Lita}}, \bibinfo {author} {\bibfnamefont {T.}~\bibnamefont {Gerrits}}, \bibinfo {author} {\bibfnamefont {S.~W.}\ \bibnamefont {Nam}}, \bibinfo {author} {\bibfnamefont {V.~D.}\ \bibnamefont {Vaidya}}, \bibinfo {author} {\bibfnamefont {M.}~\bibnamefont {Menotti}}, \bibinfo {author}
  {\bibfnamefont {I.}~\bibnamefont {Dhand}}, \bibinfo {author} {\bibfnamefont {Z.}~\bibnamefont {Vernon}}, \bibinfo {author} {\bibfnamefont {N.}~\bibnamefont {Quesada}},\ and\ \bibinfo {author} {\bibfnamefont {J.}~\bibnamefont {Lavoie}},\ }\bibfield  {title} {\bibinfo {title} {Quantum computational advantage with a programmable photonic processor},\ }\href {https://doi.org/10.1038/s41586-022-04725-x} {\bibfield  {journal} {\bibinfo  {journal} {Nature}\ }\textbf {\bibinfo {volume} {606}},\ \bibinfo {pages} {75–81} (\bibinfo {year} {2022})}\BibitemShut {NoStop}%
\bibitem [{\citenamefont {Somhorst}\ \emph {et~al.}(2023)\citenamefont {Somhorst}, \citenamefont {van~der Meer}, \citenamefont {Anguita}, \citenamefont {Schadow}, \citenamefont {Snijders}, \citenamefont {de~Goede}, \citenamefont {Kassenberg}, \citenamefont {Venderbosch}, \citenamefont {Taballione}, \citenamefont {Epping}, \citenamefont {van~den Vlekkert}, \citenamefont {Bulmer}, \citenamefont {Lugani}, \citenamefont {Walmsley}, \citenamefont {Pinkse}, \citenamefont {Eisert}, \citenamefont {Walk},\ and\ \citenamefont {Renema}}]{QuantumPhotoThermodynamics}%
  \BibitemOpen
  \bibfield  {author} {\bibinfo {author} {\bibfnamefont {F.~H.~B.}\ \bibnamefont {Somhorst}}, \bibinfo {author} {\bibfnamefont {R.}~\bibnamefont {van~der Meer}}, \bibinfo {author} {\bibfnamefont {M.~C.}\ \bibnamefont {Anguita}}, \bibinfo {author} {\bibfnamefont {R.}~\bibnamefont {Schadow}}, \bibinfo {author} {\bibfnamefont {H.~J.}\ \bibnamefont {Snijders}}, \bibinfo {author} {\bibfnamefont {M.}~\bibnamefont {de~Goede}}, \bibinfo {author} {\bibfnamefont {B.}~\bibnamefont {Kassenberg}}, \bibinfo {author} {\bibfnamefont {P.}~\bibnamefont {Venderbosch}}, \bibinfo {author} {\bibfnamefont {C.}~\bibnamefont {Taballione}}, \bibinfo {author} {\bibfnamefont {J.~P.}\ \bibnamefont {Epping}}, \bibinfo {author} {\bibfnamefont {H.~H.}\ \bibnamefont {van~den Vlekkert}}, \bibinfo {author} {\bibfnamefont {J.~F.~F.}\ \bibnamefont {Bulmer}}, \bibinfo {author} {\bibfnamefont {J.}~\bibnamefont {Lugani}}, \bibinfo {author} {\bibfnamefont {I.~A.}\ \bibnamefont {Walmsley}}, \bibinfo {author} {\bibfnamefont {P.~W.~H.}\ \bibnamefont
  {Pinkse}}, \bibinfo {author} {\bibfnamefont {J.}~\bibnamefont {Eisert}}, \bibinfo {author} {\bibfnamefont {N.}~\bibnamefont {Walk}},\ and\ \bibinfo {author} {\bibfnamefont {J.~J.}\ \bibnamefont {Renema}},\ }\bibfield  {title} {\bibinfo {title} {Quantum photo-thermodynamics on a programmable photonic quantum processor},\ }\href {https://doi.org/10.1038/s41467-023-38413-9} {\bibfield  {journal} {\bibinfo  {journal} {Nature Comm.}\ }\textbf {\bibinfo {volume} {14}},\ \bibinfo {pages} {3895} (\bibinfo {year} {2023})}\BibitemShut {NoStop}%
\bibitem [{\citenamefont {Pirandola}\ \emph {et~al.}(2015)\citenamefont {Pirandola}, \citenamefont {Eisert}, \citenamefont {Weedbrook}, \citenamefont {Furusawa},\ and\ \citenamefont {Braunstein}}]{TeleportationReview}%
  \BibitemOpen
  \bibfield  {author} {\bibinfo {author} {\bibfnamefont {S.}~\bibnamefont {Pirandola}}, \bibinfo {author} {\bibfnamefont {J.}~\bibnamefont {Eisert}}, \bibinfo {author} {\bibfnamefont {C.}~\bibnamefont {Weedbrook}}, \bibinfo {author} {\bibfnamefont {A.}~\bibnamefont {Furusawa}},\ and\ \bibinfo {author} {\bibfnamefont {S.~L.}\ \bibnamefont {Braunstein}},\ }\bibfield  {title} {\bibinfo {title} {Advances in quantum teleportation},\ }\href {https://doi.org/10.1038/nphoton.2015.154} {\bibfield  {journal} {\bibinfo  {journal} {Nature Phot.}\ }\textbf {\bibinfo {volume} {9}},\ \bibinfo {pages} {641} (\bibinfo {year} {2015})}\BibitemShut {NoStop}%
\bibitem [{\citenamefont {Burgwal}\ \emph {et~al.}(2017)\citenamefont {Burgwal}, \citenamefont {Clements}, \citenamefont {Smith}, \citenamefont {Gates}, \citenamefont {Kolthammer}, \citenamefont {Renema},\ and\ \citenamefont {Walmsley}}]{Burgwal:17}%
  \BibitemOpen
  \bibfield  {author} {\bibinfo {author} {\bibfnamefont {R.}~\bibnamefont {Burgwal}}, \bibinfo {author} {\bibfnamefont {W.~R.}\ \bibnamefont {Clements}}, \bibinfo {author} {\bibfnamefont {D.~H.}\ \bibnamefont {Smith}}, \bibinfo {author} {\bibfnamefont {J.~C.}\ \bibnamefont {Gates}}, \bibinfo {author} {\bibfnamefont {W.~S.}\ \bibnamefont {Kolthammer}}, \bibinfo {author} {\bibfnamefont {J.~J.}\ \bibnamefont {Renema}},\ and\ \bibinfo {author} {\bibfnamefont {I.~A.}\ \bibnamefont {Walmsley}},\ }\bibfield  {title} {\bibinfo {title} {Using an imperfect photonic network to implement random unitaries},\ }\href {https://doi.org/10.1364/OE.25.028236} {\bibfield  {journal} {\bibinfo  {journal} {Opt. Expr.}\ }\textbf {\bibinfo {volume} {25}},\ \bibinfo {pages} {28236} (\bibinfo {year} {2017})}\BibitemShut {NoStop}%
\bibitem [{\citenamefont {Russell}\ \emph {et~al.}(2017)\citenamefont {Russell}, \citenamefont {Chakhmakhchyan}, \citenamefont {O’Brien},\ and\ \citenamefont {Laing}}]{Russell_2017}%
  \BibitemOpen
  \bibfield  {author} {\bibinfo {author} {\bibfnamefont {N.~J.}\ \bibnamefont {Russell}}, \bibinfo {author} {\bibfnamefont {L.}~\bibnamefont {Chakhmakhchyan}}, \bibinfo {author} {\bibfnamefont {J.~L.}\ \bibnamefont {O’Brien}},\ and\ \bibinfo {author} {\bibfnamefont {A.}~\bibnamefont {Laing}},\ }\bibfield  {title} {\bibinfo {title} {Direct dialling of haar random unitary matrices},\ }\href {https://doi.org/10.1088/1367-2630/aa60ed} {\bibfield  {journal} {\bibinfo  {journal} {New J. Phys.}\ }\textbf {\bibinfo {volume} {19}},\ \bibinfo {pages} {033007} (\bibinfo {year} {2017})}\BibitemShut {NoStop}%
\bibitem [{\citenamefont {Cai}\ \emph {et~al.}(2023{\natexlab{a}})\citenamefont {Cai}, \citenamefont {Babbush}, \citenamefont {Benjamin}, \citenamefont {Endo}, \citenamefont {Huggins}, \citenamefont {Li}, \citenamefont {McClean},\ and\ \citenamefont {O'Brien}}]{RevModPhys.95.045005}%
  \BibitemOpen
  \bibfield  {author} {\bibinfo {author} {\bibfnamefont {Z.}~\bibnamefont {Cai}}, \bibinfo {author} {\bibfnamefont {R.}~\bibnamefont {Babbush}}, \bibinfo {author} {\bibfnamefont {S.~C.}\ \bibnamefont {Benjamin}}, \bibinfo {author} {\bibfnamefont {S.}~\bibnamefont {Endo}}, \bibinfo {author} {\bibfnamefont {W.~J.}\ \bibnamefont {Huggins}}, \bibinfo {author} {\bibfnamefont {Y.}~\bibnamefont {Li}}, \bibinfo {author} {\bibfnamefont {J.~R.}\ \bibnamefont {McClean}},\ and\ \bibinfo {author} {\bibfnamefont {T.~E.}\ \bibnamefont {O'Brien}},\ }\bibfield  {title} {\bibinfo {title} {Quantum error mitigation},\ }\href {https://doi.org/10.1103/RevModPhys.95.045005} {\bibfield  {journal} {\bibinfo  {journal} {Rev. Mod. Phys.}\ }\textbf {\bibinfo {volume} {95}},\ \bibinfo {pages} {045005} (\bibinfo {year} {2023}{\natexlab{a}})}\BibitemShut {NoStop}%
\bibitem [{\citenamefont {Li}\ and\ \citenamefont {Benjamin}(2017)}]{Li:2017}%
  \BibitemOpen
  \bibfield  {author} {\bibinfo {author} {\bibfnamefont {Y.}~\bibnamefont {Li}}\ and\ \bibinfo {author} {\bibfnamefont {S.~C.}\ \bibnamefont {Benjamin}},\ }\bibfield  {title} {\bibinfo {title} {{Efficient variational quantum simulator incorporating active error minimization}},\ }\href {https://doi.org/10.1103/physrevx.7.021050} {\bibfield  {journal} {\bibinfo  {journal} {Phys. Rev. X}\ }\textbf {\bibinfo {volume} {7}},\ \bibinfo {pages} {021050} (\bibinfo {year} {2017})},\ \Eprint {https://arxiv.org/abs/1611.09301} {1611.09301} \BibitemShut {NoStop}%
\bibitem [{\citenamefont {Temme}\ \emph {et~al.}(2017)\citenamefont {Temme}, \citenamefont {Bravyi},\ and\ \citenamefont {Gambetta}}]{Temme:2017}%
  \BibitemOpen
  \bibfield  {author} {\bibinfo {author} {\bibfnamefont {K.}~\bibnamefont {Temme}}, \bibinfo {author} {\bibfnamefont {S.}~\bibnamefont {Bravyi}},\ and\ \bibinfo {author} {\bibfnamefont {J.~M.}\ \bibnamefont {Gambetta}},\ }\bibfield  {title} {\bibinfo {title} {{Error mitigation for short-depth quantum circuits}},\ }\href {https://doi.org/10.1103/physrevlett.119.180509} {\bibfield  {journal} {\bibinfo  {journal} {Phys. Rev. Lett.}\ }\textbf {\bibinfo {volume} {119}},\ \bibinfo {pages} {180509} (\bibinfo {year} {2017})},\ \Eprint {https://arxiv.org/abs/1612.02058} {1612.02058} \BibitemShut {NoStop}%
\bibitem [{\citenamefont {Endo}\ \emph {et~al.}(2018)\citenamefont {Endo}, \citenamefont {Benjamin},\ and\ \citenamefont {Li}}]{Endo:2018}%
  \BibitemOpen
  \bibfield  {author} {\bibinfo {author} {\bibfnamefont {S.}~\bibnamefont {Endo}}, \bibinfo {author} {\bibfnamefont {S.~C.}\ \bibnamefont {Benjamin}},\ and\ \bibinfo {author} {\bibfnamefont {Y.}~\bibnamefont {Li}},\ }\bibfield  {title} {{\selectlanguage {English}\bibinfo {title} {{Practical quantum error mitigation for near-future applications}}},\ }\href {https://doi.org/10.1103/physrevx.8.031027} {\bibfield  {journal} {\bibinfo  {journal} {Phys. Rev. X}\ }\textbf {\bibinfo {volume} {8}},\ \bibinfo {pages} {031027} (\bibinfo {year} {2018})}\BibitemShut {NoStop}%
\bibitem [{\citenamefont {Cai}\ \emph {et~al.}(2023{\natexlab{b}})\citenamefont {Cai}, \citenamefont {Babbush}, \citenamefont {Benjamin}, \citenamefont {Endo}, \citenamefont {Huggins}, \citenamefont {Li}, \citenamefont {McClean},\ and\ \citenamefont {O’Brien}}]{Cai:2023}%
  \BibitemOpen
  \bibfield  {author} {\bibinfo {author} {\bibfnamefont {Z.}~\bibnamefont {Cai}}, \bibinfo {author} {\bibfnamefont {R.}~\bibnamefont {Babbush}}, \bibinfo {author} {\bibfnamefont {S.~C.}\ \bibnamefont {Benjamin}}, \bibinfo {author} {\bibfnamefont {S.}~\bibnamefont {Endo}}, \bibinfo {author} {\bibfnamefont {W.~J.}\ \bibnamefont {Huggins}}, \bibinfo {author} {\bibfnamefont {Y.}~\bibnamefont {Li}}, \bibinfo {author} {\bibfnamefont {J.~R.}\ \bibnamefont {McClean}},\ and\ \bibinfo {author} {\bibfnamefont {T.~E.}\ \bibnamefont {O’Brien}},\ }\bibfield  {title} {\bibinfo {title} {{Quantum error mitigation}},\ }\href {https://doi.org/10.1103/revmodphys.95.045005} {\bibfield  {journal} {\bibinfo  {journal} {Rev. Mod. Phys.}\ }\textbf {\bibinfo {volume} {95}},\ \bibinfo {pages} {045005} (\bibinfo {year} {2023}{\natexlab{b}})},\ \Eprint {https://arxiv.org/abs/2210.00921} {2210.00921} \BibitemShut {NoStop}%
\bibitem [{\citenamefont {Singh}\ \emph {et~al.}(2024)\citenamefont {Singh}, \citenamefont {Lund},\ and\ \citenamefont {Rohde}}]{PhysRevA.110.012457}%
  \BibitemOpen
  \bibfield  {author} {\bibinfo {author} {\bibfnamefont {D.}~\bibnamefont {Singh}}, \bibinfo {author} {\bibfnamefont {A.~P.}\ \bibnamefont {Lund}},\ and\ \bibinfo {author} {\bibfnamefont {P.~P.}\ \bibnamefont {Rohde}},\ }\bibfield  {title} {\bibinfo {title} {Optical cluster-state generation with unitary averaging},\ }\href {https://doi.org/10.1103/PhysRevA.110.012457} {\bibfield  {journal} {\bibinfo  {journal} {Phys. Rev. A}\ }\textbf {\bibinfo {volume} {110}},\ \bibinfo {pages} {012457} (\bibinfo {year} {2024})}\BibitemShut {NoStop}%
\bibitem [{\citenamefont {Marshman}\ \emph {et~al.}(2018)\citenamefont {Marshman}, \citenamefont {Lund}, \citenamefont {Rohde},\ and\ \citenamefont {Ralph}}]{ryan2018}%
  \BibitemOpen
  \bibfield  {author} {\bibinfo {author} {\bibfnamefont {R.~J.}\ \bibnamefont {Marshman}}, \bibinfo {author} {\bibfnamefont {A.~P.}\ \bibnamefont {Lund}}, \bibinfo {author} {\bibfnamefont {P.~P.}\ \bibnamefont {Rohde}},\ and\ \bibinfo {author} {\bibfnamefont {T.~C.}\ \bibnamefont {Ralph}},\ }\bibfield  {title} {\bibinfo {title} {Passive quantum error correction of linear optics networks through error averaging},\ }\href {https://doi.org/10.1103/PhysRevA.97.022324} {\bibfield  {journal} {\bibinfo  {journal} {Phys. Rev. A}\ }\textbf {\bibinfo {volume} {97}},\ \bibinfo {pages} {022324} (\bibinfo {year} {2018})}\BibitemShut {NoStop}%
\bibitem [{\citenamefont {Marshman}\ \emph {et~al.}(2024)\citenamefont {Marshman}, \citenamefont {Singh}, \citenamefont {Ralph},\ and\ \citenamefont {Lund}}]{ryan2024}%
  \BibitemOpen
  \bibfield  {author} {\bibinfo {author} {\bibfnamefont {R.~J.}\ \bibnamefont {Marshman}}, \bibinfo {author} {\bibfnamefont {D.}~\bibnamefont {Singh}}, \bibinfo {author} {\bibfnamefont {T.~C.}\ \bibnamefont {Ralph}},\ and\ \bibinfo {author} {\bibfnamefont {A.~P.}\ \bibnamefont {Lund}},\ }\bibfield  {title} {\bibinfo {title} {Unitary averaging with fault and loss tolerance},\ }\href {https://doi.org/10.1103/PhysRevA.109.062436} {\bibfield  {journal} {\bibinfo  {journal} {Phys. Rev. A}\ }\textbf {\bibinfo {volume} {109}},\ \bibinfo {pages} {062436} (\bibinfo {year} {2024})}\BibitemShut {NoStop}%
\bibitem [{\citenamefont {Swain}\ \emph {et~al.}(2024)\citenamefont {Swain}, \citenamefont {Marshman}, \citenamefont {Rohde}, \citenamefont {Lund}, \citenamefont {Solntsev},\ and\ \citenamefont {Ralph}}]{nibedita2024}%
  \BibitemOpen
  \bibfield  {author} {\bibinfo {author} {\bibfnamefont {S.~N.}\ \bibnamefont {Swain}}, \bibinfo {author} {\bibfnamefont {R.~J.}\ \bibnamefont {Marshman}}, \bibinfo {author} {\bibfnamefont {P.~P.}\ \bibnamefont {Rohde}}, \bibinfo {author} {\bibfnamefont {A.~P.}\ \bibnamefont {Lund}}, \bibinfo {author} {\bibfnamefont {A.~S.}\ \bibnamefont {Solntsev}},\ and\ \bibinfo {author} {\bibfnamefont {T.~C.}\ \bibnamefont {Ralph}},\ }\bibfield  {title} {\bibinfo {title} {Improving continuous-variable quantum channels with unitary averaging},\ }\href {https://doi.org/10.1103/PhysRevA.110.032622} {\bibfield  {journal} {\bibinfo  {journal} {Phys. Rev. A}\ }\textbf {\bibinfo {volume} {110}},\ \bibinfo {pages} {032622} (\bibinfo {year} {2024})}\BibitemShut {NoStop}%
\bibitem [{\citenamefont {Huggins}\ \emph {et~al.}(2021)\citenamefont {Huggins}, \citenamefont {McArdle}, \citenamefont {O'Brien}, \citenamefont {Lee}, \citenamefont {Rubin}, \citenamefont {Boixo}, \citenamefont {Whaley}, \citenamefont {Babbush},\ and\ \citenamefont {McClean}}]{PhysRevX.11.041036}%
  \BibitemOpen
  \bibfield  {author} {\bibinfo {author} {\bibfnamefont {W.~J.}\ \bibnamefont {Huggins}}, \bibinfo {author} {\bibfnamefont {S.}~\bibnamefont {McArdle}}, \bibinfo {author} {\bibfnamefont {T.~E.}\ \bibnamefont {O'Brien}}, \bibinfo {author} {\bibfnamefont {J.}~\bibnamefont {Lee}}, \bibinfo {author} {\bibfnamefont {N.~C.}\ \bibnamefont {Rubin}}, \bibinfo {author} {\bibfnamefont {S.}~\bibnamefont {Boixo}}, \bibinfo {author} {\bibfnamefont {K.~B.}\ \bibnamefont {Whaley}}, \bibinfo {author} {\bibfnamefont {R.}~\bibnamefont {Babbush}},\ and\ \bibinfo {author} {\bibfnamefont {J.~R.}\ \bibnamefont {McClean}},\ }\bibfield  {title} {\bibinfo {title} {Virtual distillation for quantum error mitigation},\ }\href {https://doi.org/10.1103/PhysRevX.11.041036} {\bibfield  {journal} {\bibinfo  {journal} {Phys. Rev. X}\ }\textbf {\bibinfo {volume} {11}},\ \bibinfo {pages} {041036} (\bibinfo {year} {2021})}\BibitemShut {NoStop}%
\bibitem [{\citenamefont {Koczor}(2021)}]{PhysRevX.11.031057}%
  \BibitemOpen
  \bibfield  {author} {\bibinfo {author} {\bibfnamefont {B.}~\bibnamefont {Koczor}},\ }\bibfield  {title} {\bibinfo {title} {Exponential error suppression for near-term quantum devices},\ }\href {https://doi.org/10.1103/PhysRevX.11.031057} {\bibfield  {journal} {\bibinfo  {journal} {Phys. Rev. X}\ }\textbf {\bibinfo {volume} {11}},\ \bibinfo {pages} {031057} (\bibinfo {year} {2021})}\BibitemShut {NoStop}%
\bibitem [{\citenamefont {Seif}\ \emph {et~al.}(2023)\citenamefont {Seif}, \citenamefont {Cian}, \citenamefont {Zhou}, \citenamefont {Chen},\ and\ \citenamefont {Jiang}}]{PRXQuantum.4.010303}%
  \BibitemOpen
  \bibfield  {author} {\bibinfo {author} {\bibfnamefont {A.}~\bibnamefont {Seif}}, \bibinfo {author} {\bibfnamefont {Z.-P.}\ \bibnamefont {Cian}}, \bibinfo {author} {\bibfnamefont {S.}~\bibnamefont {Zhou}}, \bibinfo {author} {\bibfnamefont {S.}~\bibnamefont {Chen}},\ and\ \bibinfo {author} {\bibfnamefont {L.}~\bibnamefont {Jiang}},\ }\bibfield  {title} {\bibinfo {title} {Shadow distillation: Quantum error mitigation with classical shadows for near-term quantum processors},\ }\href {https://doi.org/10.1103/PRXQuantum.4.010303} {\bibfield  {journal} {\bibinfo  {journal} {PRX Quantum}\ }\textbf {\bibinfo {volume} {4}},\ \bibinfo {pages} {010303} (\bibinfo {year} {2023})}\BibitemShut {NoStop}%
\bibitem [{\citenamefont {Onorati}\ \emph {et~al.}(2024)\citenamefont {Onorati}, \citenamefont {Kitzinger}, \citenamefont {Helsen}, \citenamefont {Ioannou}, \citenamefont {Werner}, \citenamefont {Roth},\ and\ \citenamefont {Eisert}}]{EmilioPaper}%
  \BibitemOpen
  \bibfield  {author} {\bibinfo {author} {\bibfnamefont {E.}~\bibnamefont {Onorati}}, \bibinfo {author} {\bibfnamefont {J.}~\bibnamefont {Kitzinger}}, \bibinfo {author} {\bibfnamefont {J.}~\bibnamefont {Helsen}}, \bibinfo {author} {\bibfnamefont {M.}~\bibnamefont {Ioannou}}, \bibinfo {author} {\bibfnamefont {A.~H.}\ \bibnamefont {Werner}}, \bibinfo {author} {\bibfnamefont {I.}~\bibnamefont {Roth}},\ and\ \bibinfo {author} {\bibfnamefont {J.}~\bibnamefont {Eisert}},\ }\href@noop {} {\bibinfo {title} {Noise-mitigated randomized measurements and self-calibrating shadow estimation}} (\bibinfo {year} {2024}),\ \Eprint {https://arxiv.org/abs/2403.04751} {arXiv:2403.04751} \BibitemShut {NoStop}%
\bibitem [{\citenamefont {Aaronson}\ and\ \citenamefont {Arkhipov}(2013)}]{AaransonArkhipov}%
  \BibitemOpen
  \bibfield  {author} {\bibinfo {author} {\bibfnamefont {S.}~\bibnamefont {Aaronson}}\ and\ \bibinfo {author} {\bibfnamefont {A.}~\bibnamefont {Arkhipov}},\ }\bibfield  {title} {\bibinfo {title} {The computational complexity of linear optics},\ }\href {https://doi.org/10.4086/toc.2013.v009a004} {\bibfield  {journal} {\bibinfo  {journal} {Th. Comp.}\ }\textbf {\bibinfo {volume} {9}},\ \bibinfo {pages} {143} (\bibinfo {year} {2013})}\BibitemShut {NoStop}%
\bibitem [{\citenamefont {Hangleiter}\ and\ \citenamefont {Eisert}(2023)}]{SupremacyReview}%
  \BibitemOpen
  \bibfield  {author} {\bibinfo {author} {\bibfnamefont {D.}~\bibnamefont {Hangleiter}}\ and\ \bibinfo {author} {\bibfnamefont {J.}~\bibnamefont {Eisert}},\ }\bibfield  {title} {\bibinfo {title} {Computational advantage of quantum random sampling},\ }\href {https://doi.org/10.1103/RevModPhys.95.035001} {\bibfield  {journal} {\bibinfo  {journal} {Rev. Mod. Phys.}\ }\textbf {\bibinfo {volume} {95}},\ \bibinfo {pages} {035001} (\bibinfo {year} {2023})}\BibitemShut {NoStop}%
\bibitem [{\citenamefont {Takagi}\ \emph {et~al.}(2022)\citenamefont {Takagi}, \citenamefont {Endo}, \citenamefont {Minagawa},\ and\ \citenamefont {Gu}}]{ErrorMitigationObstructionsOld}%
  \BibitemOpen
  \bibfield  {author} {\bibinfo {author} {\bibfnamefont {R.}~\bibnamefont {Takagi}}, \bibinfo {author} {\bibfnamefont {S.}~\bibnamefont {Endo}}, \bibinfo {author} {\bibfnamefont {S.}~\bibnamefont {Minagawa}},\ and\ \bibinfo {author} {\bibfnamefont {M.}~\bibnamefont {Gu}},\ }\bibfield  {title} {\bibinfo {title} {Fundamental limits of quantum error mitigation},\ }\href {https://doi.org/10.1038/s41534-022-00618-z} {\bibfield  {journal} {\bibinfo  {journal} {npj Quant. Inf.}\ }\textbf {\bibinfo {volume} {8}},\ \bibinfo {pages} {114} (\bibinfo {year} {2022})},\ \bibinfo {note} {arXiv:2210.11505}\BibitemShut {NoStop}%
\bibitem [{\citenamefont {Quek}\ \emph {et~al.}(2024)\citenamefont {Quek}, \citenamefont {França}, \citenamefont {Khatri}, \citenamefont {Meyer},\ and\ \citenamefont {Eisert}}]{Quek:2024}%
  \BibitemOpen
  \bibfield  {author} {\bibinfo {author} {\bibfnamefont {Y.}~\bibnamefont {Quek}}, \bibinfo {author} {\bibfnamefont {D.~S.}\ \bibnamefont {França}}, \bibinfo {author} {\bibfnamefont {S.}~\bibnamefont {Khatri}}, \bibinfo {author} {\bibfnamefont {J.~J.}\ \bibnamefont {Meyer}},\ and\ \bibinfo {author} {\bibfnamefont {J.}~\bibnamefont {Eisert}},\ }\bibfield  {title} {\bibinfo {title} {{Exponentially tighter bounds on limitations of quantum error mitigation}},\ }\href {https://doi.org/10.1038/s41567-024-02536-7} {\bibfield  {journal} {\bibinfo  {journal} {Nature Phys.}\ }\textbf {\bibinfo {volume} {20}},\ \bibinfo {pages} {1} (\bibinfo {year} {2024})}\BibitemShut {NoStop}%
\bibitem [{\citenamefont {Zimborás}\ \emph {et~al.}(2025)\citenamefont {Zimborás}, \citenamefont {Koczor}, \citenamefont {Holmes}, \citenamefont {Borrelli}, \citenamefont {Gilyén}, \citenamefont {Huang}, \citenamefont {Cai}, \citenamefont {Acín}, \citenamefont {Aolita}, \citenamefont {Banchi}, \citenamefont {Brandão}, \citenamefont {Cavalcanti}, \citenamefont {Cubitt}, \citenamefont {Filippov}, \citenamefont {García-Pérez}, \citenamefont {Goold}, \citenamefont {Kálmán}, \citenamefont {Kyoseva}, \citenamefont {Rossi}, \citenamefont {Sokolov}, \citenamefont {Tavernelli},\ and\ \citenamefont {Maniscalco}}]{Zimboras:2025}%
  \BibitemOpen
  \bibfield  {author} {\bibinfo {author} {\bibfnamefont {Z.}~\bibnamefont {Zimborás}}, \bibinfo {author} {\bibfnamefont {B.}~\bibnamefont {Koczor}}, \bibinfo {author} {\bibfnamefont {Z.}~\bibnamefont {Holmes}}, \bibinfo {author} {\bibfnamefont {E.-M.}\ \bibnamefont {Borrelli}}, \bibinfo {author} {\bibfnamefont {A.}~\bibnamefont {Gilyén}}, \bibinfo {author} {\bibfnamefont {H.-Y.}\ \bibnamefont {Huang}}, \bibinfo {author} {\bibfnamefont {Z.}~\bibnamefont {Cai}}, \bibinfo {author} {\bibfnamefont {A.}~\bibnamefont {Acín}}, \bibinfo {author} {\bibfnamefont {L.}~\bibnamefont {Aolita}}, \bibinfo {author} {\bibfnamefont {L.}~\bibnamefont {Banchi}}, \bibinfo {author} {\bibfnamefont {F.~G. S.~L.}\ \bibnamefont {Brandão}}, \bibinfo {author} {\bibfnamefont {D.}~\bibnamefont {Cavalcanti}}, \bibinfo {author} {\bibfnamefont {T.}~\bibnamefont {Cubitt}}, \bibinfo {author} {\bibfnamefont {S.~N.}\ \bibnamefont {Filippov}}, \bibinfo {author} {\bibfnamefont {G.}~\bibnamefont {García-Pérez}}, \bibinfo {author} {\bibfnamefont
  {J.}~\bibnamefont {Goold}}, \bibinfo {author} {\bibfnamefont {O.}~\bibnamefont {Kálmán}}, \bibinfo {author} {\bibfnamefont {E.}~\bibnamefont {Kyoseva}}, \bibinfo {author} {\bibfnamefont {M.~A.~C.}\ \bibnamefont {Rossi}}, \bibinfo {author} {\bibfnamefont {B.}~\bibnamefont {Sokolov}}, \bibinfo {author} {\bibfnamefont {I.}~\bibnamefont {Tavernelli}},\ and\ \bibinfo {author} {\bibfnamefont {S.}~\bibnamefont {Maniscalco}},\ }\href@noop {} {\bibinfo {title} {{Myths around quantum computation before full fault tolerance: What no-go theorems rule out and what they don't}}} (\bibinfo {year} {2025}),\ \Eprint {https://arxiv.org/abs/2501.05694} {arXiv:2501.05694} \BibitemShut {NoStop}%
\bibitem [{\citenamefont {Knill}\ \emph {et~al.}(2001)\citenamefont {Knill}, \citenamefont {Laflamme},\ and\ \citenamefont {Milburn}}]{knill2001klm}%
  \BibitemOpen
  \bibfield  {author} {\bibinfo {author} {\bibfnamefont {E.}~\bibnamefont {Knill}}, \bibinfo {author} {\bibfnamefont {R.}~\bibnamefont {Laflamme}},\ and\ \bibinfo {author} {\bibfnamefont {G.~J.}\ \bibnamefont {Milburn}},\ }\bibfield  {title} {\bibinfo {title} {A scheme for efficient quantum computation with linear optics},\ }\href {https://doi.org/10.1038/35051009} {\bibfield  {journal} {\bibinfo  {journal} {Nature}\ }\textbf {\bibinfo {volume} {409}},\ \bibinfo {pages} {46} (\bibinfo {year} {2001})}\BibitemShut {NoStop}%
\bibitem [{\citenamefont {Browne}\ and\ \citenamefont {Rudolph}(2005)}]{PhysRevLett.95.010501}%
  \BibitemOpen
  \bibfield  {author} {\bibinfo {author} {\bibfnamefont {D.~E.}\ \bibnamefont {Browne}}\ and\ \bibinfo {author} {\bibfnamefont {T.}~\bibnamefont {Rudolph}},\ }\bibfield  {title} {\bibinfo {title} {Resource-efficient linear optical quantum computation},\ }\href {https://doi.org/10.1103/PhysRevLett.95.010501} {\bibfield  {journal} {\bibinfo  {journal} {Phys. Rev. Lett.}\ }\textbf {\bibinfo {volume} {95}},\ \bibinfo {pages} {010501} (\bibinfo {year} {2005})}\BibitemShut {NoStop}%
\bibitem [{\citenamefont {Dawson}\ \emph {et~al.}(2006)\citenamefont {Dawson}, \citenamefont {Haselgrove},\ and\ \citenamefont {Nielsen}}]{PhysRevLett.96.020501}%
  \BibitemOpen
  \bibfield  {author} {\bibinfo {author} {\bibfnamefont {C.~M.}\ \bibnamefont {Dawson}}, \bibinfo {author} {\bibfnamefont {H.~L.}\ \bibnamefont {Haselgrove}},\ and\ \bibinfo {author} {\bibfnamefont {M.~A.}\ \bibnamefont {Nielsen}},\ }\bibfield  {title} {\bibinfo {title} {Noise thresholds for optical quantum computers},\ }\href {https://doi.org/10.1103/PhysRevLett.96.020501} {\bibfield  {journal} {\bibinfo  {journal} {Phys. Rev. Lett.}\ }\textbf {\bibinfo {volume} {96}},\ \bibinfo {pages} {020501} (\bibinfo {year} {2006})}\BibitemShut {NoStop}%
\bibitem [{\citenamefont {Gross}\ \emph {et~al.}(2006)\citenamefont {Gross}, \citenamefont {Kieling},\ and\ \citenamefont {Eisert}}]{PhysRevA.74.042343}%
  \BibitemOpen
  \bibfield  {author} {\bibinfo {author} {\bibfnamefont {D.}~\bibnamefont {Gross}}, \bibinfo {author} {\bibfnamefont {K.}~\bibnamefont {Kieling}},\ and\ \bibinfo {author} {\bibfnamefont {J.}~\bibnamefont {Eisert}},\ }\bibfield  {title} {\bibinfo {title} {Potential and limits to cluster-state quantum computing using probabilistic gates},\ }\href {https://doi.org/10.1103/PhysRevA.74.042343} {\bibfield  {journal} {\bibinfo  {journal} {Phys. Rev. A}\ }\textbf {\bibinfo {volume} {74}},\ \bibinfo {pages} {042343} (\bibinfo {year} {2006})}\BibitemShut {NoStop}%
\bibitem [{\citenamefont {Childs}\ and\ \citenamefont {Wiebe}(2012)}]{10.5555/2481569.2481570}%
  \BibitemOpen
  \bibfield  {author} {\bibinfo {author} {\bibfnamefont {A.~M.}\ \bibnamefont {Childs}}\ and\ \bibinfo {author} {\bibfnamefont {N.}~\bibnamefont {Wiebe}},\ }\bibfield  {title} {\bibinfo {title} {Hamiltonian simulation using linear combinations of unitary operations},\ }\href {https://doi.org/10.26421/QIC12.11-12} {\bibfield  {journal} {\bibinfo  {journal} {Quant. Inf. Comp.}\ }\textbf {\bibinfo {volume} {12}},\ \bibinfo {pages} {901–924} (\bibinfo {year} {2012})}\BibitemShut {NoStop}%
\bibitem [{\citenamefont {Cai}\ \emph {et~al.}(2024)\citenamefont {Cai}, \citenamefont {Tong},\ and\ \citenamefont {Preskill}}]{cai_et_al:LIPIcs.TQC.2024.2}%
  \BibitemOpen
  \bibfield  {author} {\bibinfo {author} {\bibfnamefont {Y.}~\bibnamefont {Cai}}, \bibinfo {author} {\bibfnamefont {Y.}~\bibnamefont {Tong}},\ and\ \bibinfo {author} {\bibfnamefont {J.}~\bibnamefont {Preskill}},\ }\bibfield  {title} {\bibinfo {title} {{Stochastic error cancellation in analog quantum simulation}},\ }in\ \href {https://doi.org/10.4230/LIPIcs.TQC.2024.2} {\emph {\bibinfo {booktitle} {19th Conference on the Theory of Quantum Computation, Communication and Cryptography (TQC 2024)}}},\ \bibinfo {series} {Leibniz International Proceedings in Informatics (LIPIcs)}, Vol.\ \bibinfo {volume} {310},\ \bibinfo {editor} {edited by\ \bibinfo {editor} {\bibfnamefont {F.}~\bibnamefont {Magniez}}\ and\ \bibinfo {editor} {\bibfnamefont {A.~B.}\ \bibnamefont {Grilo}}}\ (\bibinfo  {publisher} {Schloss Dagstuhl -- Leibniz-Zentrum f{\"u}r Informatik},\ \bibinfo {address} {Dagstuhl, Germany},\ \bibinfo {year} {2024})\ pp.\ \bibinfo {pages} {2:1--2:15}\BibitemShut {NoStop}%
\bibitem [{\citenamefont {Nielsen}\ and\ \citenamefont {Chuang}(2010)}]{NielsenMichaelA}%
  \BibitemOpen
  \bibfield  {author} {\bibinfo {author} {\bibfnamefont {M.~A.}\ \bibnamefont {Nielsen}}\ and\ \bibinfo {author} {\bibfnamefont {I.~L.}\ \bibnamefont {Chuang}},\ }\href@noop {} {\emph {\bibinfo {title} {Quantum computation and quantum information}}},\ \bibinfo {edition} {10th}\ ed.\ (\bibinfo  {publisher} {Cambridge University Press},\ \bibinfo {address} {Cambridge},\ \bibinfo {year} {2010})\BibitemShut {NoStop}%
\bibitem [{\citenamefont {Arkhipov}(2015)}]{Arkhipov_2015}%
  \BibitemOpen
  \bibfield  {author} {\bibinfo {author} {\bibfnamefont {A.}~\bibnamefont {Arkhipov}},\ }\bibfield  {title} {\bibinfo {title} {{BosonSampling} is robust against small errors in the network matrix},\ }\href {https://doi.org/10.1103/physreva.92.062326} {\bibfield  {journal} {\bibinfo  {journal} {Phys. Rev. A}\ }\textbf {\bibinfo {volume} {92}},\ \bibinfo {pages} {062326} (\bibinfo {year} {2015})}\BibitemShut {NoStop}%
\bibitem [{\citenamefont {Reck}\ \emph {et~al.}(1994)\citenamefont {Reck}, \citenamefont {Zeilinger}, \citenamefont {Bernstein},\ and\ \citenamefont {Bertani}}]{Reck}%
  \BibitemOpen
  \bibfield  {author} {\bibinfo {author} {\bibfnamefont {M.}~\bibnamefont {Reck}}, \bibinfo {author} {\bibfnamefont {A.}~\bibnamefont {Zeilinger}}, \bibinfo {author} {\bibfnamefont {H.~J.}\ \bibnamefont {Bernstein}},\ and\ \bibinfo {author} {\bibfnamefont {P.}~\bibnamefont {Bertani}},\ }\bibfield  {title} {\bibinfo {title} {Experimental realization of any discrete unitary operator},\ }\href {https://doi.org/10.1103/PhysRevLett.73.58} {\bibfield  {journal} {\bibinfo  {journal} {Phys. Rev. Lett.}\ }\textbf {\bibinfo {volume} {73}},\ \bibinfo {pages} {58} (\bibinfo {year} {1994})}\BibitemShut {NoStop}%
\bibitem [{\citenamefont {Clements}\ \emph {et~al.}(2016)\citenamefont {Clements}, \citenamefont {Humphreys}, \citenamefont {Metcalf}, \citenamefont {Kolthammer},\ and\ \citenamefont {Walmsley}}]{Clements}%
  \BibitemOpen
  \bibfield  {author} {\bibinfo {author} {\bibfnamefont {W.~R.}\ \bibnamefont {Clements}}, \bibinfo {author} {\bibfnamefont {P.~C.}\ \bibnamefont {Humphreys}}, \bibinfo {author} {\bibfnamefont {B.~J.}\ \bibnamefont {Metcalf}}, \bibinfo {author} {\bibfnamefont {W.~S.}\ \bibnamefont {Kolthammer}},\ and\ \bibinfo {author} {\bibfnamefont {I.~A.}\ \bibnamefont {Walmsley}},\ }\bibfield  {title} {\bibinfo {title} {Optimal design for universal multiport interferometers},\ }\href {https://doi.org/10.1364/OPTICA.3.001460} {\bibfield  {journal} {\bibinfo  {journal} {Optica}\ }\textbf {\bibinfo {volume} {3}},\ \bibinfo {pages} {1460} (\bibinfo {year} {2016})}\BibitemShut {NoStop}%
\bibitem [{\citenamefont {Wassner}\ \emph {et~al.}(2025)\citenamefont {Wassner}, \citenamefont {Guaita}, \citenamefont {Eisert},\ and\ \citenamefont {Carrasco}}]{ClaraPaper}%
  \BibitemOpen
  \bibfield  {author} {\bibinfo {author} {\bibfnamefont {C.}~\bibnamefont {Wassner}}, \bibinfo {author} {\bibfnamefont {T.}~\bibnamefont {Guaita}}, \bibinfo {author} {\bibfnamefont {J.}~\bibnamefont {Eisert}},\ and\ \bibinfo {author} {\bibfnamefont {J.}~\bibnamefont {Carrasco}},\ }\href@noop {} {\bibinfo {title} {Holonomic quantum computation: A scalable adiabatic architecture}} (\bibinfo {year} {2025}),\ \Eprint {https://arxiv.org/abs/2502.17188} {arXiv:2502.17188} \BibitemShut {NoStop}%
\bibitem [{\citenamefont {Lund}(2023)}]{Lund2023}%
  \BibitemOpen
  \bibfield  {author} {\bibinfo {author} {\bibfnamefont {A.~P.}\ \bibnamefont {Lund}},\ }\bibfield  {title} {\bibinfo {title} {{Estimating Fock-state linear optics evolution using coherent states}},\ }\href {https://doi.org/10.1116/5.0136828} {\bibfield  {journal} {\bibinfo  {journal} {AVS Quant. Sc.}\ }\textbf {\bibinfo {volume} {5}},\ \bibinfo {pages} {011405} (\bibinfo {year} {2023})}\BibitemShut {NoStop}%
\bibitem [{\citenamefont {Alexander}\ \emph {et~al.}(2025)\citenamefont {Alexander}, \citenamefont {Benyamini}, \citenamefont {Black}, \citenamefont {Bonneau}, \citenamefont {Burgos}, \citenamefont {Burridge}, \citenamefont {Cable}, \citenamefont {Campbell}, \citenamefont {Catalano}, \citenamefont {Ceballos}, \citenamefont {Chang}, \citenamefont {Choudhury}, \citenamefont {Chung}, \citenamefont {Danesh}, \citenamefont {Dauer}, \citenamefont {Davis}, \citenamefont {Dudley}, \citenamefont {Er-Xuan}, \citenamefont {Fargas}, \citenamefont {Farsi}, \citenamefont {Fenrich}, \citenamefont {Frazer}, \citenamefont {Fukami}, \citenamefont {Ganesan}, \citenamefont {Gibson}, \citenamefont {Gimeno-Segovia}, \citenamefont {Goeldi}, \citenamefont {Goley}, \citenamefont {Haislmaier}, \citenamefont {Halimi}, \citenamefont {Hansen}, \citenamefont {Hardy}, \citenamefont {Horng}, \citenamefont {House}, \citenamefont {Hu}, \citenamefont {Jadidi}, \citenamefont {Jain}, \citenamefont {Johansson}, \citenamefont {Jones},
  \citenamefont {Kamineni}, \citenamefont {Kelez}, \citenamefont {Koustuban}, \citenamefont {Kovall}, \citenamefont {Krogen}, \citenamefont {Kumar}, \citenamefont {Liang}, \citenamefont {LiCausi}, \citenamefont {Llewellyn}, \citenamefont {Lokovic}, \citenamefont {Lovelady}, \citenamefont {Manfrinato}, \citenamefont {Melnichuk}, \citenamefont {Mendoza}, \citenamefont {Moores}, \citenamefont {Mukherjee}, \citenamefont {Munns}, \citenamefont {Musalem}, \citenamefont {Najafi}, \citenamefont {O’Brien}, \citenamefont {Ortmann}, \citenamefont {Pai}, \citenamefont {Park}, \citenamefont {Peng}, \citenamefont {Penthorn}, \citenamefont {Peterson}, \citenamefont {Peterson}, \citenamefont {Poush}, \citenamefont {Pryde}, \citenamefont {Ramprasad}, \citenamefont {Ray}, \citenamefont {Rodriguez}, \citenamefont {Roxworthy}, \citenamefont {Rudolph}, \citenamefont {Saunders}, \citenamefont {Shadbolt}, \citenamefont {Shah}, \citenamefont {Bahgat~Shehata}, \citenamefont {Shin}, \citenamefont {Sinsky}, \citenamefont {Smith},
  \citenamefont {Sohn}, \citenamefont {Sohn}, \citenamefont {Son}, \citenamefont {Souza}, \citenamefont {Sparrow}, \citenamefont {Staffaroni}, \citenamefont {Stavrakas}, \citenamefont {Sukumaran}, \citenamefont {Tamborini}, \citenamefont {Thompson}, \citenamefont {Tran}, \citenamefont {Triplett}, \citenamefont {Tung}, \citenamefont {Veitia}, \citenamefont {Vert}, \citenamefont {Vidrighin}, \citenamefont {Vorobeichik}, \citenamefont {Weigel}, \citenamefont {Wingert}, \citenamefont {Wooding}, \citenamefont {Zhou},\ and\ \citenamefont {{PsiQuantum Team}}}]{alexander_manufacturable_2025}%
  \BibitemOpen
  \bibfield  {author} {\bibinfo {author} {\bibfnamefont {K.}~\bibnamefont {Alexander}}, \bibinfo {author} {\bibfnamefont {A.}~\bibnamefont {Benyamini}}, \bibinfo {author} {\bibfnamefont {D.}~\bibnamefont {Black}}, \bibinfo {author} {\bibfnamefont {D.}~\bibnamefont {Bonneau}}, \bibinfo {author} {\bibfnamefont {S.}~\bibnamefont {Burgos}}, \bibinfo {author} {\bibfnamefont {B.}~\bibnamefont {Burridge}}, \bibinfo {author} {\bibfnamefont {H.}~\bibnamefont {Cable}}, \bibinfo {author} {\bibfnamefont {G.}~\bibnamefont {Campbell}}, \bibinfo {author} {\bibfnamefont {G.}~\bibnamefont {Catalano}}, \bibinfo {author} {\bibfnamefont {A.}~\bibnamefont {Ceballos}}, \bibinfo {author} {\bibfnamefont {C.-M.}\ \bibnamefont {Chang}}, \bibinfo {author} {\bibfnamefont {S.~S.}\ \bibnamefont {Choudhury}}, \bibinfo {author} {\bibfnamefont {C.}~\bibnamefont {Chung}}, \bibinfo {author} {\bibfnamefont {F.}~\bibnamefont {Danesh}}, \bibinfo {author} {\bibfnamefont {T.}~\bibnamefont {Dauer}}, \bibinfo {author} {\bibfnamefont {M.}~\bibnamefont
  {Davis}}, \bibinfo {author} {\bibfnamefont {E.}~\bibnamefont {Dudley}}, \bibinfo {author} {\bibfnamefont {P.}~\bibnamefont {Er-Xuan}}, \bibinfo {author} {\bibfnamefont {J.}~\bibnamefont {Fargas}}, \bibinfo {author} {\bibfnamefont {A.}~\bibnamefont {Farsi}}, \bibinfo {author} {\bibfnamefont {C.}~\bibnamefont {Fenrich}}, \bibinfo {author} {\bibfnamefont {J.}~\bibnamefont {Frazer}}, \bibinfo {author} {\bibfnamefont {M.}~\bibnamefont {Fukami}}, \bibinfo {author} {\bibfnamefont {Y.}~\bibnamefont {Ganesan}}, \bibinfo {author} {\bibfnamefont {G.}~\bibnamefont {Gibson}}, \bibinfo {author} {\bibfnamefont {M.}~\bibnamefont {Gimeno-Segovia}}, \bibinfo {author} {\bibfnamefont {S.}~\bibnamefont {Goeldi}}, \bibinfo {author} {\bibfnamefont {P.}~\bibnamefont {Goley}}, \bibinfo {author} {\bibfnamefont {R.}~\bibnamefont {Haislmaier}}, \bibinfo {author} {\bibfnamefont {S.}~\bibnamefont {Halimi}}, \bibinfo {author} {\bibfnamefont {P.}~\bibnamefont {Hansen}}, \bibinfo {author} {\bibfnamefont {S.}~\bibnamefont {Hardy}}, \bibinfo
  {author} {\bibfnamefont {J.}~\bibnamefont {Horng}}, \bibinfo {author} {\bibfnamefont {M.}~\bibnamefont {House}}, \bibinfo {author} {\bibfnamefont {H.}~\bibnamefont {Hu}}, \bibinfo {author} {\bibfnamefont {M.}~\bibnamefont {Jadidi}}, \bibinfo {author} {\bibfnamefont {V.}~\bibnamefont {Jain}}, \bibinfo {author} {\bibfnamefont {H.}~\bibnamefont {Johansson}}, \bibinfo {author} {\bibfnamefont {T.}~\bibnamefont {Jones}}, \bibinfo {author} {\bibfnamefont {V.}~\bibnamefont {Kamineni}}, \bibinfo {author} {\bibfnamefont {N.}~\bibnamefont {Kelez}}, \bibinfo {author} {\bibfnamefont {R.}~\bibnamefont {Koustuban}}, \bibinfo {author} {\bibfnamefont {G.}~\bibnamefont {Kovall}}, \bibinfo {author} {\bibfnamefont {P.}~\bibnamefont {Krogen}}, \bibinfo {author} {\bibfnamefont {N.}~\bibnamefont {Kumar}}, \bibinfo {author} {\bibfnamefont {Y.}~\bibnamefont {Liang}}, \bibinfo {author} {\bibfnamefont {N.}~\bibnamefont {LiCausi}}, \bibinfo {author} {\bibfnamefont {D.}~\bibnamefont {Llewellyn}}, \bibinfo {author} {\bibfnamefont
  {K.}~\bibnamefont {Lokovic}}, \bibinfo {author} {\bibfnamefont {M.}~\bibnamefont {Lovelady}}, \bibinfo {author} {\bibfnamefont {V.~R.}\ \bibnamefont {Manfrinato}}, \bibinfo {author} {\bibfnamefont {A.}~\bibnamefont {Melnichuk}}, \bibinfo {author} {\bibfnamefont {G.}~\bibnamefont {Mendoza}}, \bibinfo {author} {\bibfnamefont {B.}~\bibnamefont {Moores}}, \bibinfo {author} {\bibfnamefont {S.}~\bibnamefont {Mukherjee}}, \bibinfo {author} {\bibfnamefont {J.}~\bibnamefont {Munns}}, \bibinfo {author} {\bibfnamefont {F.-X.}\ \bibnamefont {Musalem}}, \bibinfo {author} {\bibfnamefont {F.}~\bibnamefont {Najafi}}, \bibinfo {author} {\bibfnamefont {J.~L.}\ \bibnamefont {O’Brien}}, \bibinfo {author} {\bibfnamefont {J.~E.}\ \bibnamefont {Ortmann}}, \bibinfo {author} {\bibfnamefont {S.}~\bibnamefont {Pai}}, \bibinfo {author} {\bibfnamefont {B.}~\bibnamefont {Park}}, \bibinfo {author} {\bibfnamefont {H.-T.}\ \bibnamefont {Peng}}, \bibinfo {author} {\bibfnamefont {N.}~\bibnamefont {Penthorn}}, \bibinfo {author}
  {\bibfnamefont {B.}~\bibnamefont {Peterson}}, \bibinfo {author} {\bibfnamefont {G.}~\bibnamefont {Peterson}}, \bibinfo {author} {\bibfnamefont {M.}~\bibnamefont {Poush}}, \bibinfo {author} {\bibfnamefont {G.~J.}\ \bibnamefont {Pryde}}, \bibinfo {author} {\bibfnamefont {T.}~\bibnamefont {Ramprasad}}, \bibinfo {author} {\bibfnamefont {G.}~\bibnamefont {Ray}}, \bibinfo {author} {\bibfnamefont {A.~V.}\ \bibnamefont {Rodriguez}}, \bibinfo {author} {\bibfnamefont {B.}~\bibnamefont {Roxworthy}}, \bibinfo {author} {\bibfnamefont {T.}~\bibnamefont {Rudolph}}, \bibinfo {author} {\bibfnamefont {D.~J.}\ \bibnamefont {Saunders}}, \bibinfo {author} {\bibfnamefont {P.}~\bibnamefont {Shadbolt}}, \bibinfo {author} {\bibfnamefont {D.}~\bibnamefont {Shah}}, \bibinfo {author} {\bibfnamefont {A.}~\bibnamefont {Bahgat~Shehata}}, \bibinfo {author} {\bibfnamefont {H.}~\bibnamefont {Shin}}, \bibinfo {author} {\bibfnamefont {J.}~\bibnamefont {Sinsky}}, \bibinfo {author} {\bibfnamefont {J.}~\bibnamefont {Smith}}, \bibinfo {author}
  {\bibfnamefont {B.}~\bibnamefont {Sohn}}, \bibinfo {author} {\bibfnamefont {Y.-I.}\ \bibnamefont {Sohn}}, \bibinfo {author} {\bibfnamefont {G.}~\bibnamefont {Son}}, \bibinfo {author} {\bibfnamefont {M.~C. M.~M.}\ \bibnamefont {Souza}}, \bibinfo {author} {\bibfnamefont {C.}~\bibnamefont {Sparrow}}, \bibinfo {author} {\bibfnamefont {M.}~\bibnamefont {Staffaroni}}, \bibinfo {author} {\bibfnamefont {C.}~\bibnamefont {Stavrakas}}, \bibinfo {author} {\bibfnamefont {V.}~\bibnamefont {Sukumaran}}, \bibinfo {author} {\bibfnamefont {D.}~\bibnamefont {Tamborini}}, \bibinfo {author} {\bibfnamefont {M.~G.}\ \bibnamefont {Thompson}}, \bibinfo {author} {\bibfnamefont {K.}~\bibnamefont {Tran}}, \bibinfo {author} {\bibfnamefont {M.}~\bibnamefont {Triplett}}, \bibinfo {author} {\bibfnamefont {M.}~\bibnamefont {Tung}}, \bibinfo {author} {\bibfnamefont {A.}~\bibnamefont {Veitia}}, \bibinfo {author} {\bibfnamefont {A.}~\bibnamefont {Vert}}, \bibinfo {author} {\bibfnamefont {M.~D.}\ \bibnamefont {Vidrighin}}, \bibinfo {author}
  {\bibfnamefont {I.}~\bibnamefont {Vorobeichik}}, \bibinfo {author} {\bibfnamefont {P.}~\bibnamefont {Weigel}}, \bibinfo {author} {\bibfnamefont {M.}~\bibnamefont {Wingert}}, \bibinfo {author} {\bibfnamefont {J.}~\bibnamefont {Wooding}}, \bibinfo {author} {\bibfnamefont {X.}~\bibnamefont {Zhou}},\ and\ \bibinfo {author} {\bibnamefont {{PsiQuantum Team}}},\ }\bibfield  {title} {\bibinfo {title} {A manufacturable platform for photonic quantum computing},\ }\bibfield  {journal} {\bibinfo  {journal} {Nature}\ }\href {https://doi.org/10.1038/s41586-025-08820-7} {10.1038/s41586-025-08820-7} (\bibinfo {year} {2025})\BibitemShut {NoStop}%
\bibitem [{\citenamefont {Somhorst}\ \emph {et~al.}(2025)\citenamefont {Somhorst}, \citenamefont {Sau\"er}, \citenamefont {van~den Hoven},\ and\ \citenamefont {Renema}}]{PhysRevApplied.23.044003}%
  \BibitemOpen
  \bibfield  {author} {\bibinfo {author} {\bibfnamefont {F.}~\bibnamefont {Somhorst}}, \bibinfo {author} {\bibfnamefont {B.}~\bibnamefont {Sau\"er}}, \bibinfo {author} {\bibfnamefont {S.}~\bibnamefont {van~den Hoven}},\ and\ \bibinfo {author} {\bibfnamefont {J.}~\bibnamefont {Renema}},\ }\bibfield  {title} {\bibinfo {title} {Photon-distillation schemes with reduced resource costs based on multiphoton fourier interference},\ }\href {https://doi.org/10.1103/PhysRevApplied.23.044003} {\bibfield  {journal} {\bibinfo  {journal} {Phys. Rev. Appl.}\ }\textbf {\bibinfo {volume} {23}},\ \bibinfo {pages} {044003} (\bibinfo {year} {2025})}\BibitemShut {NoStop}%
\bibitem [{\citenamefont {Ralph}\ \emph {et~al.}(2005)\citenamefont {Ralph}, \citenamefont {Hayes},\ and\ \citenamefont {Gilchrist}}]{PhysRevLett.95.100501}%
  \BibitemOpen
  \bibfield  {author} {\bibinfo {author} {\bibfnamefont {T.~C.}\ \bibnamefont {Ralph}}, \bibinfo {author} {\bibfnamefont {A.~J.~F.}\ \bibnamefont {Hayes}},\ and\ \bibinfo {author} {\bibfnamefont {A.}~\bibnamefont {Gilchrist}},\ }\bibfield  {title} {\bibinfo {title} {Loss-tolerant optical qubits},\ }\href {https://doi.org/10.1103/PhysRevLett.95.100501} {\bibfield  {journal} {\bibinfo  {journal} {Phys. Rev. Lett.}\ }\textbf {\bibinfo {volume} {95}},\ \bibinfo {pages} {100501} (\bibinfo {year} {2005})}\BibitemShut {NoStop}%
\bibitem [{\citenamefont {Hayes}\ \emph {et~al.}(2008)\citenamefont {Hayes}, \citenamefont {Gilchrist},\ and\ \citenamefont {Ralph}}]{PhysRevA.77.012310}%
  \BibitemOpen
  \bibfield  {author} {\bibinfo {author} {\bibfnamefont {A.~J.~F.}\ \bibnamefont {Hayes}}, \bibinfo {author} {\bibfnamefont {A.}~\bibnamefont {Gilchrist}},\ and\ \bibinfo {author} {\bibfnamefont {T.~C.}\ \bibnamefont {Ralph}},\ }\bibfield  {title} {\bibinfo {title} {Loss-tolerant operations in parity-code linear optics quantum computing},\ }\href {https://doi.org/10.1103/PhysRevA.77.012310} {\bibfield  {journal} {\bibinfo  {journal} {Phys. Rev. A}\ }\textbf {\bibinfo {volume} {77}},\ \bibinfo {pages} {012310} (\bibinfo {year} {2008})}\BibitemShut {NoStop}%
\bibitem [{\citenamefont {Ryser}(1963)}]{ryser}%
  \BibitemOpen
  \bibfield  {author} {\bibinfo {author} {\bibfnamefont {H.~J.}\ \bibnamefont {Ryser}},\ }\href {http://www.jstor.org/stable/10.4169/j.ctt5hh8v6} {\emph {\bibinfo {title} {Combinatorial Mathematics}}},\ \bibinfo {edition} {1st}\ ed.,\ Vol.~\bibinfo {volume} {14}\ (\bibinfo  {publisher} {Mathematical Association of America},\ \bibinfo {year} {1963})\BibitemShut {NoStop}%
\bibitem [{\citenamefont {Seron}\ \emph {et~al.}(2024)\citenamefont {Seron}, \citenamefont {Novo}, \citenamefont {Arkhipov},\ and\ \citenamefont {Cerf}}]{seron2022efficientvalidationbosonsampling}%
  \BibitemOpen
  \bibfield  {author} {\bibinfo {author} {\bibfnamefont {B.}~\bibnamefont {Seron}}, \bibinfo {author} {\bibfnamefont {L.}~\bibnamefont {Novo}}, \bibinfo {author} {\bibfnamefont {A.}~\bibnamefont {Arkhipov}},\ and\ \bibinfo {author} {\bibfnamefont {N.~J.}\ \bibnamefont {Cerf}},\ }\bibfield  {title} {\bibinfo {title} {Efficient validation of boson sampling from binned photon-number distributions},\ }\href {https://doi.org/10.22331/q-2024-09-19-1479} {\bibfield  {journal} {\bibinfo  {journal} {Quantum}\ }\textbf {\bibinfo {volume} {8}},\ \bibinfo {pages} {1479} (\bibinfo {year} {2024})}\BibitemShut {NoStop}%
\bibitem [{\citenamefont {Schlimgen}\ \emph {et~al.}(2021)\citenamefont {Schlimgen}, \citenamefont {Head-Marsden}, \citenamefont {Sager}, \citenamefont {Narang},\ and\ \citenamefont {Mazziotti}}]{PhysRevLett.127.270503}%
  \BibitemOpen
  \bibfield  {author} {\bibinfo {author} {\bibfnamefont {A.~W.}\ \bibnamefont {Schlimgen}}, \bibinfo {author} {\bibfnamefont {K.}~\bibnamefont {Head-Marsden}}, \bibinfo {author} {\bibfnamefont {L.~M.}\ \bibnamefont {Sager}}, \bibinfo {author} {\bibfnamefont {P.}~\bibnamefont {Narang}},\ and\ \bibinfo {author} {\bibfnamefont {D.~A.}\ \bibnamefont {Mazziotti}},\ }\bibfield  {title} {\bibinfo {title} {Quantum simulation of open quantum systems using a unitary decomposition of operators},\ }\href {https://doi.org/10.1103/PhysRevLett.127.270503} {\bibfield  {journal} {\bibinfo  {journal} {Phys. Rev. Lett.}\ }\textbf {\bibinfo {volume} {127}},\ \bibinfo {pages} {270503} (\bibinfo {year} {2021})}\BibitemShut {NoStop}%
\bibitem [{\citenamefont {Gily\'{e}n}\ \emph {et~al.}(2019)\citenamefont {Gily\'{e}n}, \citenamefont {Su}, \citenamefont {Low},\ and\ \citenamefont {Wiebe}}]{10.1145/3313276.3316366}%
  \BibitemOpen
  \bibfield  {author} {\bibinfo {author} {\bibfnamefont {A.}~\bibnamefont {Gily\'{e}n}}, \bibinfo {author} {\bibfnamefont {Y.}~\bibnamefont {Su}}, \bibinfo {author} {\bibfnamefont {G.~H.}\ \bibnamefont {Low}},\ and\ \bibinfo {author} {\bibfnamefont {N.}~\bibnamefont {Wiebe}},\ }\bibfield  {title} {\bibinfo {title} {Quantum singular value transformation and beyond: exponential improvements for quantum matrix arithmetics},\ }in\ \href {https://doi.org/10.1145/3313276.3316366} {\emph {\bibinfo {booktitle} {Proceedings of the 51st Annual ACM SIGACT Symposium on Theory of Computing}}},\ \bibinfo {series and number} {STOC 2019}\ (\bibinfo  {publisher} {Association for Computing Machinery},\ \bibinfo {address} {New York, NY, USA},\ \bibinfo {year} {2019})\ p.\ \bibinfo {pages} {193–204}\BibitemShut {NoStop}%
\bibitem [{\citenamefont {Saghaei}\ \emph {et~al.}(2019)\citenamefont {Saghaei}, \citenamefont {Elyasi},\ and\ \citenamefont {Karimzadeh}}]{SAGHAEI2019100733}%
  \BibitemOpen
  \bibfield  {author} {\bibinfo {author} {\bibfnamefont {H.}~\bibnamefont {Saghaei}}, \bibinfo {author} {\bibfnamefont {P.}~\bibnamefont {Elyasi}},\ and\ \bibinfo {author} {\bibfnamefont {R.}~\bibnamefont {Karimzadeh}},\ }\bibfield  {title} {\bibinfo {title} {{Design, fabrication, and characterization of Mach–Zehnder interferometers}},\ }\href {https://doi.org/https://doi.org/10.1016/j.photonics.2019.100733} {\bibfield  {journal} {\bibinfo  {journal} {Phot. Nanostruct. Fund. Appl.}\ }\textbf {\bibinfo {volume} {37}},\ \bibinfo {pages} {100733} (\bibinfo {year} {2019})}\BibitemShut {NoStop}%
\bibitem [{\citenamefont {Rahimi-Keshari}\ \emph {et~al.}(2013)\citenamefont {Rahimi-Keshari}, \citenamefont {Broome}, \citenamefont {Fickler}, \citenamefont {Fedrizzi}, \citenamefont {Ralph},\ and\ \citenamefont {White}}]{Rahimi-Keshari:13}%
  \BibitemOpen
  \bibfield  {author} {\bibinfo {author} {\bibfnamefont {S.}~\bibnamefont {Rahimi-Keshari}}, \bibinfo {author} {\bibfnamefont {M.~A.}\ \bibnamefont {Broome}}, \bibinfo {author} {\bibfnamefont {R.}~\bibnamefont {Fickler}}, \bibinfo {author} {\bibfnamefont {A.}~\bibnamefont {Fedrizzi}}, \bibinfo {author} {\bibfnamefont {T.~C.}\ \bibnamefont {Ralph}},\ and\ \bibinfo {author} {\bibfnamefont {A.~G.}\ \bibnamefont {White}},\ }\bibfield  {title} {\bibinfo {title} {Direct characterization of linear-optical networks},\ }\href {https://doi.org/10.1364/OE.21.013450} {\bibfield  {journal} {\bibinfo  {journal} {Opt. Express}\ }\textbf {\bibinfo {volume} {21}},\ \bibinfo {pages} {13450} (\bibinfo {year} {2013})}\BibitemShut {NoStop}%
\bibitem [{\citenamefont {Laing}\ and\ \citenamefont {O'Brien}(2012)}]{Laing:2012}%
  \BibitemOpen
  \bibfield  {author} {\bibinfo {author} {\bibfnamefont {A.}~\bibnamefont {Laing}}\ and\ \bibinfo {author} {\bibfnamefont {J.~L.}\ \bibnamefont {O'Brien}},\ }\href {https://doi.org/10.48550/arxiv.1208.2868} {\bibinfo {title} {Super-stable tomography of any linear optical device}} (\bibinfo {year} {2012}),\ \Eprint {https://arxiv.org/abs/1208.2868} {arXiv:1208.2868} \BibitemShut {NoStop}%
\bibitem [{\citenamefont {Hoch}\ \emph {et~al.}(2023)\citenamefont {Hoch}, \citenamefont {Giordani}, \citenamefont {Spagnolo}, \citenamefont {Crespi}, \citenamefont {Osellame},\ and\ \citenamefont {Sciarrino}}]{Hoch:2023}%
  \BibitemOpen
  \bibfield  {author} {\bibinfo {author} {\bibfnamefont {F.}~\bibnamefont {Hoch}}, \bibinfo {author} {\bibfnamefont {T.}~\bibnamefont {Giordani}}, \bibinfo {author} {\bibfnamefont {N.}~\bibnamefont {Spagnolo}}, \bibinfo {author} {\bibfnamefont {A.}~\bibnamefont {Crespi}}, \bibinfo {author} {\bibfnamefont {R.}~\bibnamefont {Osellame}},\ and\ \bibinfo {author} {\bibfnamefont {F.}~\bibnamefont {Sciarrino}},\ }\bibfield  {title} {\bibinfo {title} {{Characterization of multimode linear optical networks}},\ }\href {https://doi.org/10.1117/1.apn.2.1.016007} {\bibfield  {journal} {\bibinfo  {journal} {Adv. Phot. Nex.}\ }\textbf {\bibinfo {volume} {2}},\ \bibinfo {pages} {016007} (\bibinfo {year} {2023})},\ \Eprint {https://arxiv.org/abs/2304.06486} {2304.06486} \BibitemShut {NoStop}%
\bibitem [{\citenamefont {Fyrillas}\ \emph {et~al.}(2024)\citenamefont {Fyrillas}, \citenamefont {Faure}, \citenamefont {Maring}, \citenamefont {Senellart},\ and\ \citenamefont {Belabas}}]{Fyrillas:2024}%
  \BibitemOpen
  \bibfield  {author} {\bibinfo {author} {\bibfnamefont {A.}~\bibnamefont {Fyrillas}}, \bibinfo {author} {\bibfnamefont {O.}~\bibnamefont {Faure}}, \bibinfo {author} {\bibfnamefont {N.}~\bibnamefont {Maring}}, \bibinfo {author} {\bibfnamefont {J.}~\bibnamefont {Senellart}},\ and\ \bibinfo {author} {\bibfnamefont {N.}~\bibnamefont {Belabas}},\ }\bibfield  {title} {\bibinfo {title} {{Scalable machine learning-assisted clear-box characterization for optimally controlled photonic circuits}},\ }\href {https://doi.org/10.1364/optica.512148} {\bibfield  {journal} {\bibinfo  {journal} {Optica}\ }\textbf {\bibinfo {volume} {11}},\ \bibinfo {pages} {427} (\bibinfo {year} {2024})}\BibitemShut {NoStop}%
\bibitem [{\citenamefont {Park}\ \emph {et~al.}(2023)\citenamefont {Park}, \citenamefont {Matsumoto}, \citenamefont {Kiyohara}, \citenamefont {Hofmann}, \citenamefont {Okamoto},\ and\ \citenamefont {Takeuchi}}]{Park2023}%
  \BibitemOpen
  \bibfield  {author} {\bibinfo {author} {\bibfnamefont {G.}~\bibnamefont {Park}}, \bibinfo {author} {\bibfnamefont {I.}~\bibnamefont {Matsumoto}}, \bibinfo {author} {\bibfnamefont {T.}~\bibnamefont {Kiyohara}}, \bibinfo {author} {\bibfnamefont {H.~F.}\ \bibnamefont {Hofmann}}, \bibinfo {author} {\bibfnamefont {R.}~\bibnamefont {Okamoto}},\ and\ \bibinfo {author} {\bibfnamefont {S.}~\bibnamefont {Takeuchi}},\ }\bibfield  {title} {\bibinfo {title} {Realization of photon correlations beyond the linear optics limit},\ }\href {https://doi.org/10.1126/sciadv.adj8146} {\bibfield  {journal} {\bibinfo  {journal} {Science Adv.}\ }\textbf {\bibinfo {volume} {9}},\ \bibinfo {pages} {eadj8146} (\bibinfo {year} {2023})}\BibitemShut {NoStop}%
\end{thebibliography}%

\end{document}